\documentclass[twocolumn]{aastex61}

\newcommand\aastex{AAS\TeX}

\received{March 21, 2018}
\accepted{July 18, 2018}
\submitjournal{ApJ}

\shorttitle{\aastex\ sample article}
\shortauthors{Kadowaki et al.}
\watermark{DRAFT}
\usepackage{amsmath}
\usepackage{epstopdf}

\begin{document}

\title{MHD instabilities in accretion disks and their implications in driving fast magnetic reconnection}

\correspondingauthor{Luis H.S. Kadowaki}
\email{luis.kadowaki@iag.usp.br, dalpino@iag.usp.br}

\author[0000-0002-6908-5634]{Luis H.S. Kadowaki}
\altaffiliation{FAPESP Fellowship}
\affil{Universidade de S\~{a}o Paulo, Instituto de Astronomia, Geof\'{i}sica e Ci\^{e}ncias Atmosf\'{e}ricas, Departamento de Astronomia \\
       1226 Mat\~{a}o Street \\
       S\~{a}o Paulo, 05508-090, Brasil}

\author{Elisabete M. de Gouveia Dal Pino}
\affiliation{Universidade de S\~{a}o Paulo, Instituto de Astronomia, Geof\'{i}sica e Ci\^{e}ncias Atmosf\'{e}ricas, Departamento de Astronomia \\
             1226 Mat\~{a}o Street \\
             S\~{a}o Paulo, 05508-090, Brasil}

\author{James M. Stone}
\affiliation{Department of Astrophysical Sciences, Peyton Hall, Princeton University \\
             Princeton, NJ 08544, USA}



\begin{abstract}
Magnetohydrodynamic instabilities play an important role in accretion disks systems.
Besides the well-known effects of the magnetorotational instability (MRI), the Parker-Rayleigh-Taylor instability (PRTI) also arises as an important mechanism to help in the formation of the coronal region around an accretion disk and in the production of magnetic reconnection events similar to those occurring in the solar corona.
In this work, we have performed three-dimensional magnetohydrodynamical (3D-MHD) shearing-box numerical simulations of accretion disks with an initial stratified density distribution and a strong azimuthal magnetic field with a ratio between the thermal and magnetic pressures of the order of unity. This study aimed at verifying the role of these instabilities \added{in driving fast magnetic reconnection  in turbulent accretion disk/corona systems}. \added{As we expected, the simulations showed} an initial formation of large-scale magnetic loops due to the PRTI followed by the development of a nearly steady-state turbulence driven by both instabilities. \added{In this turbulent environment,} we have employed an algorithm to identify the presence of current sheets produced by the encounter of magnetic flux ropes of opposite polarity in the turbulent regions of both the corona and the disk. We computed the magnetic reconnection rates in these locations obtaining average reconnection velocities in Alfv\'en speed units of the order of \added{$0.13 \pm 0.09$} in the accretion disk and $0.17 \pm 0.10 $ in the coronal region (with mean peak values of order of $0.2$), which are consistent with the predictions of  the theory of turbulence-induced fast reconnection. 
\end{abstract}




\section{Introduction}
\label{sec:intro}

\added{Accretion disk systems are ubiquitous in Astrophysical environments at all scales \citep[for reviews see, e.g.,][]{pringle_81, balbus_hawley_98, abramowski_fragile_13}.} The complex emission features observed from these systems indicate the existence of different regimes of accretion and frequently a hot, low-density magnetized disk corona is invoked in order to explain non-thermal high-energy emission, such as the well known X-ray transitions observed in black hole binaries
\citep[BHBs, see, e.g.,][]{fender_etal_04, belloni_etal_05, remillard_mcclintock_06, kylafis_belloni_15}. These transitions are characterized by a Low/Hard state attributed to the inverse Compton process in a geometrically thick optically thin accretion flow at a sub-Eddington regime \citep[see][]{esin_etal_97, esin_etal_98, esin_etal_01, narayan_mcclintock_08}, also known as ADAF \citep[advection-dominated accretion flow, see][]{narayan_yi_94, narayan_yi_95, abramowski_etal_95}. These systems also show a High/Soft state generally attributed to the heating of a geometrically thin, optically thick accretion disk \citep{shakura_sunyaev_73} at near Eddington regime and, between these two states, there is a transient one  \citep[of the order of a few days; see][]{remillard_mcclintock_06}, whose origin is not fully understood yet, but could be explained by shocks in a jet \citep[e.g.,][]{romero_etal_03, bosch-ramon_etal_05,piano_etal_12} or reconnection in the corona \citep[e.g.,][]{dgdp_lazarian_05, dgdp_etal_10b,dgdp_etal_10a, kadowaki_etal_15}. 

Magnetic reconnection in accretion disk/corona systems as a mean to explain flaring emission has been  explored in several analytical and numerical works \citep[see, e.g.,][]{dgdp_lazarian_05, igumenshchev_09, soker_10, uzdensky_spitkovsky_14, dexter_etal_14, huang_etal_14, kadowaki_etal_15, singh_etal_15, khiali_etal_15a}. For instance, \citet{dgdp_lazarian_05} proposed an analytical model to explain the peculiar flaring state of the BHB (also named microquasar) GRS$1915$+$105$, which was later  extended to other few BHBs, \added{active galactic nuclei} (AGNs) and protostars \citep[see][]{dgdp_etal_10b,dgdp_etal_10a}. According to this model, fast reconnection between the magnetic field lines of the source's magnetosphere and those raising from  the accretion disk can produce plasmoids and accelerate particles to relativistic velocities via a first-order Fermi process in the current sheets \citep[see][and references therein]{dgdp_lazarian_05, drake_etal_06, drake_etal_10, zenitani_hoshino_08, zenitani_etal_09, kowal_etal_11, kowal_etal_12, delvalle_etal_16,guo_etal_16}, providing enough power to explain observed non-thermal flaring emission. More recently, employing a similar model, but with the  fast reconnection driven by the background turbulence in the coronal plasma permeating the magnetic field lines,  \cite{kadowaki_etal_15} and \cite{singh_etal_15} verified that this could explain the very high-energy emission of hundreds of low-luminosity (non-Blazar) AGNs and BHBs. A similar process has been proposed for  energy extraction in the vicinity of Kerr black holes by \cite{koide_arai_08}. 
Magnetic reconnection of the field lines generated by buoyancy processes like the Parker-Rayleigh-Taylor instability \citep[PRTI, see][]{parker_66} has been also invoked to be an efficient process to heat the coronal region of accretion disks around black holes by \citet[]{liu_etal_02, liu_etal_03} and \citet{huang_etal_14}.

The efficiency of the reconnection in such cases is a key ingredient to allow for highly variable, explosive  emission, as in the case of the solar flares which show magnetic reconnection velocities in a range between $(0.001-0.5) V_{A}$, where $V_{A}$ is the Alfv\'en speed \cite[see, e.g.,][]{dere_1996, aschwanden_etal_01, su_etal_2013}. 

Several mechanisms have been invoked  to describe  magnetic reconnection. Since the proposed Sweet-Parker model \citep{sweet_58,parker_57} that predicts very slow reconnection rates:

\begin{equation}
V_{rec}=V_{in}/V_{A} = S^{-1/2}~,
\end{equation}
where $V_{in}$ is the reconnection velocity of opposite magnetic fluxes, $S=L V_{A} / \eta$ is the Lundquist number, $L$ the length of reconnection site, and $\eta$ the Ohmic resistivity, other models have been proposed in order to explain observed fast reconnection, like the Petschek's X-point model. In this model, reconnection is forced to occur in single points, rather than in an entire flat current sheet \citep[with a reconnection rate $V_{rec} \propto (\ln{S})^{-1}$, see][]{petschek_64}.
This increases the reconnection rate, but \cite{biskamp_etal_97} showed numerically that this mechanism is unstable and makes the system evolve rapidly to a Sweet-Parker configuration, unless the flow is collisionless and has localized variable resistivity. The main difficulties in explaining  fast reconnection in collisional flows have been removed with the proposal of the model of  \cite{lazarian_vishiniac_99} that  considers  the  ubiquitous presence of turbulence in astrophysical environments to drive fast reconnection. The idea behind this model is that the wandering of the magnetic field lines in a turbulent flow allows for  several patches to reconnect simultaneously making reconnection fast \citep[][]{lazarian_vishiniac_99}:

\begin{equation}
V_{rec} = \min \left[ \left( \frac{L}{l}\right)^{1/2}, \left( \frac{l}{L}\right)^{1/2} \right] \left( \frac{V_{l}}{V_{A}} \right)^{2}~,
\label{eq:vrec_turb}
\end{equation}
where $V_{l}$ and $l$ are the velocity and size of the turbulent eddies at injection scale, and independent of the background resistivity, so that even nearly ideal MHD flows that have nearly negligible Ohmic resistivity, as most astrophysical flows, can undergo fast reconnection when there is turbulence. Extensive 3D numerical MHD testing of this model has been successfully performed in current sheets with embedded forced turbulence \citep[e.g.,][]{kowal_etal_09, kowal_etal_12, takamoto_etal_15}, but it has never been thoroughly investigated in systems with natural driving sources of turbulence, like the magneto-rotational instability \citep[MRI,][]{chandrasekhar_60, balbus_hawley_91, hawley_etal_95, balbus_hawley_98} and the PRTI in accretion disks \citep[see, however, tentative exploration of fast reconnection driven by current-kink instability turbulence in 3D MHD relativistic jet simulations by][]{singh_etal_16}.

\added{The role of MHD instabilities in the dynamics of accretion disks has been extensively studied in several works. Most of these studies have explored numerically the evolution of the MRI considering either a zero net magnetic field flux \citep[see][]{brandenburg_etal_95, davis_etal_10, miller_stone_2000, simon_etal_2011, simon_etal_2012}, or a net vertical field flux \citep{hawley_etal_95, bai_stone_13, salvesen_etal_16b}. 
Besides the MRI, other magnetohydrodynamic (MHD) processes can play an important role in accretion disks, for instance, in the building and evolution of a hot, low-density magnetized corona around them. In particular, the PRTI is an important mechanism to drive, under some conditions, the formation of magnetic loops arising from the disk out of an initially horizontal magnetic field \citep[see,][]{parker_55,parker_66,isobe_etal_05}.
In recent years, \cite{johansen_levin_08} and \cite{salvesen_etal_16a} have demonstrated the importance of the vertical transport of the magnetic field from the disk to the outer parts of the system due to the PRTI in strongly magnetized regimes (which can actually be found in AGNs and BHBs). However, the magnetic reconnection process and its efficiency have not been explored in these works.}  

Motivated by the studies described above that have highlighted  the potential importance of  turbulence and fast magnetic reconnection  in  magnetized  accretion disk coronal flows to explain  observed phenomena including  particle acceleration  and non-thermal flare emission features in compact sources, 
our  main goal  here is to explore these processes  quantitatively in depth, which require very high resolution numerical simulations. For this reason, we have performed  local 3D-MHD numerical simulations employing a shearing-box approximation \citep[][]{hawley_etal_95}  \added{considering an initial  stratified density distribution and}  strong horizontal magnetic fields. We could assess the role of the PRTI and MRI in the development of turbulence, large scale  magnetic loops in the corona,  and fast magnetic reconnection driven by turbulence. We have also used a modified algorithm based on the work of \cite{zhdankin_etal_13} and \cite{kowal_etal_09} to identify reconnection sites and evaluate statistically the magnetic reconnection rates in the accretion disk and corona. 

The paper is organized as follows, in section \ref{sec:num_met} we describe the numerical method for the shearing-box approach \citep{hawley_etal_95} and the initial and boundary conditions used in this work \citep[see, e.g.,][]{johansen_levin_08, davis_etal_10, bai_stone_13}. In section \ref{sec:num_results} we show the results obtained from the evolution of the averages of the physical quantities taken over  the accretion disk and coronal regions.
\added{In section \ref{sec:mag_rec} we show the results of the identification of fast reconnection driven by turbulence in the disk and corona, as well as the statistical properties of these sites and the corresponding reconnection rates. Finally, in section \ref{sec:discussion_conclusion} we discuss the relevant physical proprieties of our simulations, compare our results with previous works when pertinent \citep[see, e.g.,][]{miller_stone_2000, johansen_levin_08,salvesen_etal_16b,salvesen_etal_16a}, then highlight the main results found on magnetic reconnection, and draw our conclusions.}

\section{Numerical method}
\label{sec:num_met}

The numerical solutions of our simulations are obtained from the magnetohydrodynamics equations (MHD) which describe the macroscopic behavior of a magnetized fluid. These equations, in the conservative and ideal form, are given by:

\begin{equation}
\label{eq.mass}
\frac{\partial\rho}{\partial t}+\nabla.(\rho {\boldsymbol{v}})=0 ~~,
\end{equation}
\begin{equation}
\label{eq.momentum}
\begin{split}
\frac{\partial\rho {\boldsymbol{v}}}{\partial t} &+ \nabla. \left[\rho {\boldsymbol{vv}} + \left(P+ \frac{{\boldsymbol{B}}.{\boldsymbol{B}}}{8\pi}\right) {\boldsymbol{I}} - \frac{{\boldsymbol{BB}}}{4\pi}\right] \\
& = \rho \left[2 {\boldsymbol{v}} \times {\boldsymbol{\Omega}} + 2q\Omega_{0}^{2}x\widehat{\boldsymbol{x}} - \Omega_{0}^{2}z\widehat{\boldsymbol{z}}  \right] ~~,
\end{split}
\end{equation}
\begin{equation}
\label{eq.induction}
\frac{\partial {\boldsymbol{B}}}{\partial t}+\nabla \times({\boldsymbol{B}} \times {\boldsymbol{v}})=0~~,
\end{equation}
and correspond to the equations of mass and momentum conservation, and induction, respectively. In these equations, $\rho$ is the density, $P$ is the thermal pressure of the gas, $\boldsymbol{v}$ is the velocity field, ${\boldsymbol{I}}$ is the identity matrix, and $\boldsymbol{B}$ is the magnetic field. The equations have been solved in nondimensional form, thus without a factor $4\pi$. In the present work, we adopt an isothermal equation of state with $P=c_{s}^2\rho$, where $c_{s}$ is the sound speed.

\added{We employ}  a shearing-box approach \citep[see][]{hawley_etal_95}, useful to obtain the statistical properties of very small regions of accretion disks systems. Such approach consists in a linear shearing velocity in the azimuthal direction ``$y$'' given by:

\begin{equation}
\label{linear_shear}
v_{y}=-q\Omega_{0}x
~~\textrm{with} ~~ q=-\frac{d\ln{\Omega(R)}}{d\ln{R}}\bigg\vert_{R=R_{0}} 
\end{equation}
where $\Omega_{0}$ is the angular velocity at an arbitrary radius $R_{0}$, and $q=3/2$ is the shear parameter for a Keplerian profile. 

\added{From this approach (in the disk reference frame at $R_{0}$), the source term of the eq.(\ref{eq.momentum}) on the right hand side ``$2 \rho {\boldsymbol{v}} \times {\boldsymbol{\Omega}}$'' (for $\boldsymbol{\Omega}=\Omega_{0}\widehat{\boldsymbol{z}}$) corresponds to the Coriolis force, whereas ``$2\rho q\Omega_{0}^{2}x~\widehat{\boldsymbol{x}}$'' corresponds to the  effective forces (centrifugal $+$ gravitational) obtained from the local expansion of the momentum equation \citep[see, e.g.,][]{balbus_hawley_98}, and ``$- \rho\Omega_{0}^{2}z~\widehat{\boldsymbol{z}}$'' corresponds to the vertical gravity term due to the vertical density stratification.}

We have used the ATHENA code to obtain the numerical solution of the 3D-MHD equations with the shearing-box approach \citep[see][]{stone_etal_08,stone_etal_10, stone_gardiner_10}. To compute the intercell fluxes of the computational grid, a HLLD Riemann solver has been employed \citep[see][]{miyoshi_kusano_05}, while a second-order Runge-Kutta scheme has been used to solve the equations in time. An orbital advection scheme has also been adopted \citep{stone_gardiner_10}, where the azimuthal component of the velocity ($v_{y}$) is split into an advection part ($-3/2\Omega_{0} x$) and another one involving only fluctuations ($u_{y}$) given by:

\begin{equation}
\label{eq:fargo}
\boldsymbol{u}=\boldsymbol{v}+\frac{3}{2}\Omega_{0} x \widehat{\boldsymbol{y}} ~~.
\end{equation}
This scheme improves the simulations, since the numerical integration of the advection part is not subject to the Courant-Friedrich-Lewy (CFL) condition, making the time step ($dt$) less restrictive \added{\citep[see,][]{masset_00, johnson_etal_2008, davis_etal_10, stone_gardiner_10}}.

\subsection{Initial conditions}

As initial conditions, we have used a density profile obtained from a magnetostatic equilibrium in the vertical direction since we have adopted a strong azimuthal magnetic field to trigger the PRTI. The density profile is given by \citep[see][]{johansen_levin_08}:

\begin{equation}
\label{eq:rho_profile}
\rho(z)=\rho_{0}\exp[-z^2/2H_{\beta}^2] ~~,
\end{equation}
where $\rho_{0}=1$ is the initial density in midplane of the accretion disk, $H_{\beta}=\sqrt{1+\beta_{0}^{-1}} c_{s}/\Omega_{0}$ is the scale height of the gas, and $\beta_{0}$ is the initial ratio between the thermal and magnetic pressures. For consistency with the work of \cite{johansen_levin_08}, we have adopted $\Omega_{0}=1$, and the thermal scale height as our unit of length ($H=c_{s}/\Omega_{0} = 1$), that yields to $c_{s}=1$. Assuming a constant $\beta_{0}$, the magnetic field profile is given by \citep[see also][]{johansen_levin_08}:

\begin{equation}
\label{eq:mag_profile}
B_{y}=\sqrt{\rho_{0}\exp[-z^2/2H_{\beta}^2]\Omega_{0}^2/\beta_{0}} ~~,
\end{equation}
where we have taken $B_{x}=B_{z}=0$. It is important to emphasize that the criteria to trigger the PRTI will be slightly different compared with the well-known work of \citet{parker_66}, since we are dealing with a differential rotation system under a linear gravity. The presence of a differential rotation was studied analytically by \citet{foglizzo_tagger_94, foglizzo_tagger_95}, where they obtained new instability conditions and studied the correlations between the PRTI and MRI. \citet{kim_etal_97}, on the other hand, studied the instability criteria under a linear gravity, more appropriate to the case of Keplerian disks and the present work. However, in all the works, an initial $\beta_{0}$ of the order of the unit is still required to trigger the PRTI.

We have assumed the initial velocity field as zero ($\boldsymbol{u}=0$), since an orbital advection scheme has also been adopted \citep[see eq.\ref{eq:fargo} and][]{stone_gardiner_10}. However, a Gaussian perturbation with an amplitude of $\delta u \sim 10^{-3}$  has been used in the azimuthal velocity component to trigger the linear phase of the PRTI and MRI \citep[see][]{johansen_levin_08}.

\subsection{Boundary conditions}

We have used shearing-periodic boundary conditions in the radial direction ``$x$'' that reproduce the differential rotation through the azimuthal displacement of the radial boundaries \citep[see][]{hawley_etal_95}. In the azimuthal direction ``$y$'' we have used standard periodic boundary conditions, and outflow boundaries in the vertical direction ``$z$'' which are more appropriate to the study and evolution of the coronal region in the surroundings of accretion disks \citep[see][]{bai_stone_13}. In outflow boundary conditions, zero-gradients are used for the velocity and magnetic fields, but the vertical velocity component is set zero ($v_{z}=0$) whether inflows are detected through the boundaries. Besides, the density is extrapolated assuming a vertical hydrostatic equilibrium. 

Finally, a density floor ($\rho_{min}$) has been applied to minimize \added{the numerical limitations due to the high ``physical'' Alfv\'{e}n speed} \citep[see, e.g.,][]{bai_stone_13} since the rarefied corona is subject to the action of a strong magnetic field (that is transported by the magnetic buoyancy) with the ratio between the thermal and magnetic pressure of the order of unit. \added{We notice that this floor could change the physics of the problem  for breaking down the initial magnetostatic equilibrium. However, we verified along the simulation that the horizontally averaged densities are larger than the adopted density floor \citep[as in][]{bai_stone_13}, so  that this procedure has a minimum impact on the evolution of the system.} We have also used a mass diffusion term in the mass conservation equation \citep[see, e.g.,][]{gressel_etal_2011} to alleviate the density floor condition\footnote{Different to the work of \citet{gressel_etal_2011}, our implementation does not conserve mass and for this reason, a density floor has been used.} and allow us to use $\rho_{min} = 10^{-6}$ \added{for the low-resolution, $\rho_{min} = 10^{-6}$, $10^{-7}$, or $10^{-8}$ for the intermediate resolution,} and $\rho_{min} = 10^{-5}$ for the high-resolution. \added{For two of the models with intermediate resolution (see next subsection for more details), due to the weak magnetic field intensity applied as initial condition and the magnetostatic equilibrium adjust, the density profile in the vertical direction has been found to decay faster than in the other models for floor values $10^{-6}$ or $10^{-5}$, thus affecting strongly the initial evolution of the system, as mentioned above. In these cases we  adopted floors values of $10^{-7}$ and $10^{-8}$ which are lower than the horizontally averaged density at the high altitudes in these models.}  

\subsection{Simulation parameters}

In this work, we have used resolutions of $\sim 11$ ($128 \times 128 \times 128$), $\sim 21$ ($256 \times 256 \times 256$), and $\sim 43$ ($512 \times 512 \times 512$) cells per thermal height scale of the system ($H = c_s \Omega^{-1}$), considering a computational domain of size $12H \times 12H \times 12H$. We also considered three different values of $\beta_{0}$ ($1$, $10$, and $100$) to evaluate the evolution of \added{the MRI and PRTI} instabilities \added{and turbulence} in the system. 
 The parameters of the simulations are shown in Table \ref{tab:sb} and each model name is composed by the resolution (R11, R21, and R43) plus the initial value of $\beta_{0}$ (b1, b10, and b100). The diagnostics used to analyze these models are presented in the next section.

\begin{center}
\scriptsize
\begin{longtable*}{lccccccc}
\caption[Simulation parameters]{Simulation parameters.}\\
\hline \hline \\[-2ex]
\multicolumn{1}{c}{Simulation} &
\multicolumn{1}{c}{Computacional domain} &
\multicolumn{1}{c}{Resolution} &
\multicolumn{1}{c}{$\beta_{0}$} &
\multicolumn{1}{c}{$\Omega$} &
\multicolumn{1}{c}{$q$} &
\multicolumn{1}{c}{$\rho_{min}$} &
\multicolumn{1}{c}{Vertical boundaries}

\\[0.5ex] \hline
\\[-1.8ex]

\endfirsthead

\multicolumn{8}{c}{\footnotesize{{\slshape{{\tablename} \thetable{}}} - Continued}}\\[0.5ex]

\hline \hline\\[-2ex]

\multicolumn{1}{c}{Simulation} &
\multicolumn{1}{c}{Computacional domain} &
\multicolumn{1}{c}{Resolution} &
\multicolumn{1}{c}{$\beta_{0}$} &
\multicolumn{1}{c}{$\Omega$} &
\multicolumn{1}{c}{$q$} &
\multicolumn{1}{c}{$\rho_{min}$} &
\multicolumn{1}{c}{Vertical boundaries}

\\[0.5ex] \hline
\\[-1.8ex]

\endhead

\multicolumn{3}{l}{{\footnotesize{Continue in the next page\ldots}}}\\
\endfoot
\hline

\endlastfoot

R11b1 &$12H\times 12H\times 12H$ &$128\times 128\times 128$ &$1.0 $ &$1.0$ &$1.5$ & $10^{-6}$ & \textit{Outflows} \\
%PMRIg\_11H\_oxyz12\_b0010
%M11\_b10 &$12H\times 12H\times 12H$ &$128\times 128\times 128$ &$10  $ &$1.0$ &$1.5$ &\textit{Outflows} \\
%PMRIg\_11H\_oxyz12\_b0100
%M11\_b100 &$12H\times 12H\times 12H$ &$128\times 128\times 128$ &$10^2$ &$1.0$ &$1.5$ &\textit{Outflows} \\
%PMRIg\_11H\_pxyz12          &$12H\times 12H\times 12H$ &$128\times 128\times 128$ &$1.0 $ &$1.0$ &$1.5$ &Periodic        \\
%PMRIg\_21H\_oxyz12
R21b1 &$12H\times 12H\times 12H$ &$256\times 256\times 256$ &$1.0 $ &$1.0$ &$1.5$ & $10^{-6}$ & \textit{Outflows} \\
%PMRIg\_21H\_oxyz12\_b0010
R21b10 &$12H\times 12H\times 12H$ &$256\times 256\times 256$ &$10  $ &$1.0$ &$1.5$ & $10^{-7}$ & \textit{Outflows} \\
%PMRIg\_21H\_oxyz12\_b0100
R21b100 &$12H\times 12H\times 12H$ &$256\times 256\times 256$ &$10^2$ &$1.0$ &$1.5$ & $10^{-8}$ & \textit{Outflows} \\
%PMRIg\_43H\_oxyz12
R43b1 &$12H\times 12H\times 12H$ &$512\times 512\times 512$ &$1.0 $ &$1.0$ &$1.5$ & $10^{-5}$ & \textit{Outflows} \\

\label{tab:sb}
\end{longtable*}
\end{center}

\subsection{Diagnostics}
\label{sec:Diagnostics}

In order to follow the evolution of the system and the instabilities, we will present here  time and space average diagrams of relevant quantities. The space average of a given variable $f$ taken at the entire volume of the box is represented as $\left< f \right>$, so that:

\begin{equation}
\left<f\right>=\frac{\int f dxdydz}{\int dxdy dz} ~~.
\end{equation}

Also, in order to track the evolution of $f$ at the vertical direction ``$z$'', we performed space averages at ``$xy$'' plane, at each height ``$z$'' (horizontal averages), represented by the symbol $\left <f \right>_{xy}$, where:

\begin{equation}
\left< f \right>_{xy}=\frac{\int f dxdy}{\int dxdy} ~~.
\end{equation}

Finally, time averages are indicated by a $t$ index, so that time volume and horizontal averages are represented by $\left< f \right>_{t}$ and $\left< f \right>_{xyt}$, respectively.

\added{In this work, we have evaluated the angular momentum transport parameter $\alpha$ \citep{shakura_sunyaev_73} through the relation:}

\begin{equation}
\alpha \equiv \frac{T_{xy}}{P} = \frac{T_{xy}^{Max}+T_{xy}^{Rey}}{\rho c_{s}^2} ~~,
\label{eq:alpha}
\end{equation}
\added{where $T_{xy}^{Max}=-B_{x}B_{y}$ is the Maxwell (magnetic) stress tensor, and $T_{xy}^{Rey}=\rho u_{x}u_{y}$ is the Reynolds stress tensor. We have also evaluated the ratio between the thermal and magnetic pressures, given by:}

\begin{equation}
\beta = \frac{\rho c_{s}^2}{B^2/2} ~~.
\label{eq:beta}
\end{equation}
\added{The volume and horizontal averages of eqs.(\ref{eq:alpha}) and (\ref{eq:beta}) are represented by an upper bar (i.e., $\left<\overline{\alpha}\right>$, $\left<\overline{\alpha}\right>_{xy}$,  $\left<\overline{\beta}\right>$, and $\left<\overline{\beta}\right>_{xy}$) to indicate the ratio of the averages of two different quantities, for example:}

\begin{equation}
\left<\overline{\alpha}\right> = \frac{\left<T_{xy}\right>}{\left<\rho c_{s}^2\right>} ~~ \textrm{and} ~~ \left<\overline{\alpha}\right>_{xy} = \frac{\left<T_{xy}\right>_{xy}}{\left<\rho c_{s}^2\right>_{xy}}
\label{eq:barnotation}
\end{equation}

To evaluate the power spectrum of the velocity field ($|\boldsymbol{\widetilde{u}}(k_x, k_y,z)|^2$), we  performed a two-dimensional Fourier transformation to obtain the $k_x$ and $k_y$ wavenumbers in each height ``$z$''  of the domain:

\begin{equation}
\begin{split}
\boldsymbol{\widetilde{u}}(k_x, k_y,z) &= {1\over{\sqrt{N_x N_y}}} \\
& \times \sum_{n,m=0}^{N_x-1, N_y-1} \boldsymbol{u}(x_n, y_m,z) e^{-2\pi i({k_x x_n\over N_{x}}+{k_y y_m\over N_{y}})},
\label{eq:power_spectrum}
\end{split}
\end{equation}
where $N_{x}$ and $N_{y}$ are the total number of cells in the radial and azimuthal directions, respectively. The spectrum $P(k,z)$ has been obtained by integrating $|\boldsymbol{\widetilde{u}}(k_x, k_y,z)|^2$ over \added{annular areas between $k-dk$ and $k$, where $k=\sqrt(k_x^2+k_y^2)$}.

We adopt the R21b1 model (see table \ref{tab:sb}) as our reference one. We describe the main results of this simulation in the next section.

\section{Numerical results}
\label{sec:num_results}

\subsection{R21b1 model}

\begin{figure*}
    \centering
	\includegraphics[scale=0.265]{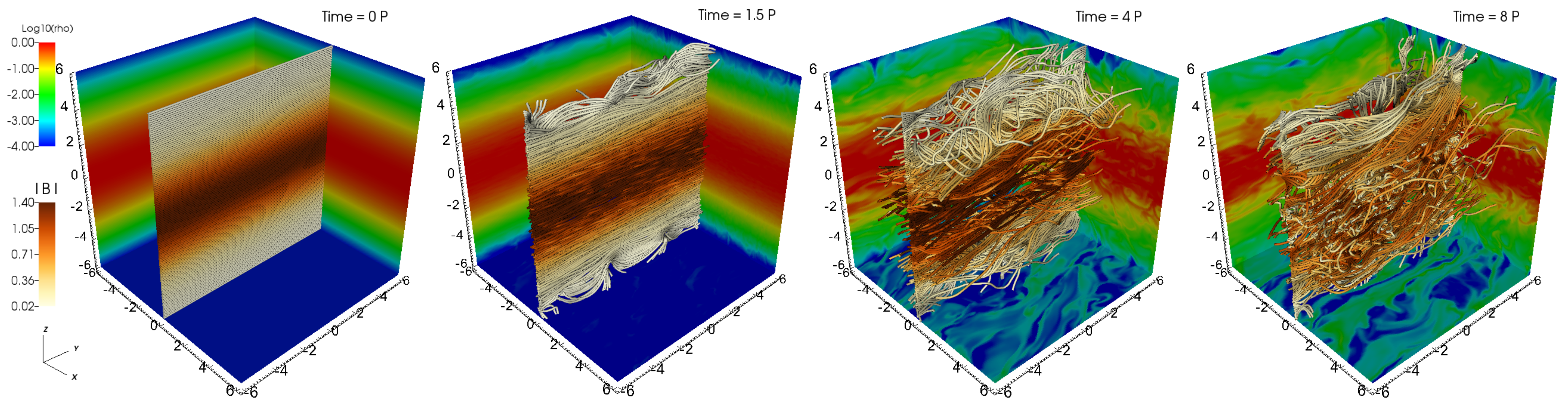}
	\caption{The diagrams show the time evolution of the total intensity of the magnetic field (streamlines) and the density distribution (background colors in a logarithmic scale) of the model R21b1 between $t=0$ and $t=8P$, highlighting the central slice.} 
	\label{fig:21H_ic}
\end{figure*}

 Figure \ref{fig:21H_ic} shows the time evolution of the magnetic field (streamlines) and the density distribution (background colors in a logarithmic scale) at $t=0$, $1.5P$, $4P$, and $8P$ (where $P=2\pi\Omega_{0}^{-1}$ corresponds to \added{one} orbital period of the accretion disk), for the R21b1 model (see Table \ref{tab:sb}). The configuration of the magnetic field lines at $t=1.5P$ shows that the azimuthal component ``$B_y$'' is stretched in the vertical direction ``$z$'' due to the effects of magnetic buoyancy, producing loops during the exponential growth of the PRTI \citep[which are in agreement with the results obtained by][]{johansen_levin_08}. 
The development of the turbulence, the decrease of the intensity of the magnetic field and its transport from the midplane of the disk to the coronal region due to the PRTI (and to outside the computational domain) are observed at $t=4P$ and $8P$.

\begin{figure} 
	\includegraphics[scale=0.45]{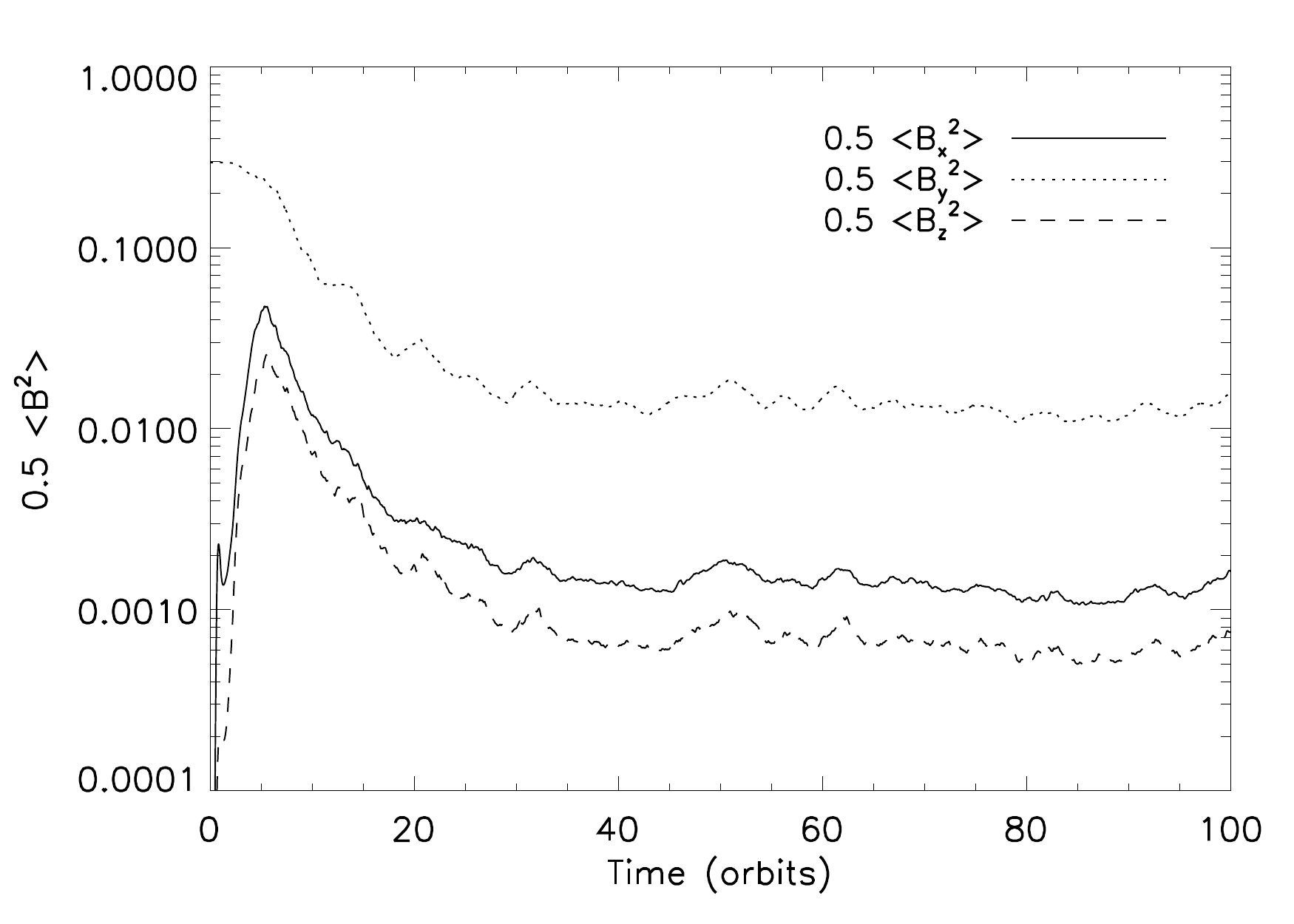}
	\includegraphics[scale=0.45]{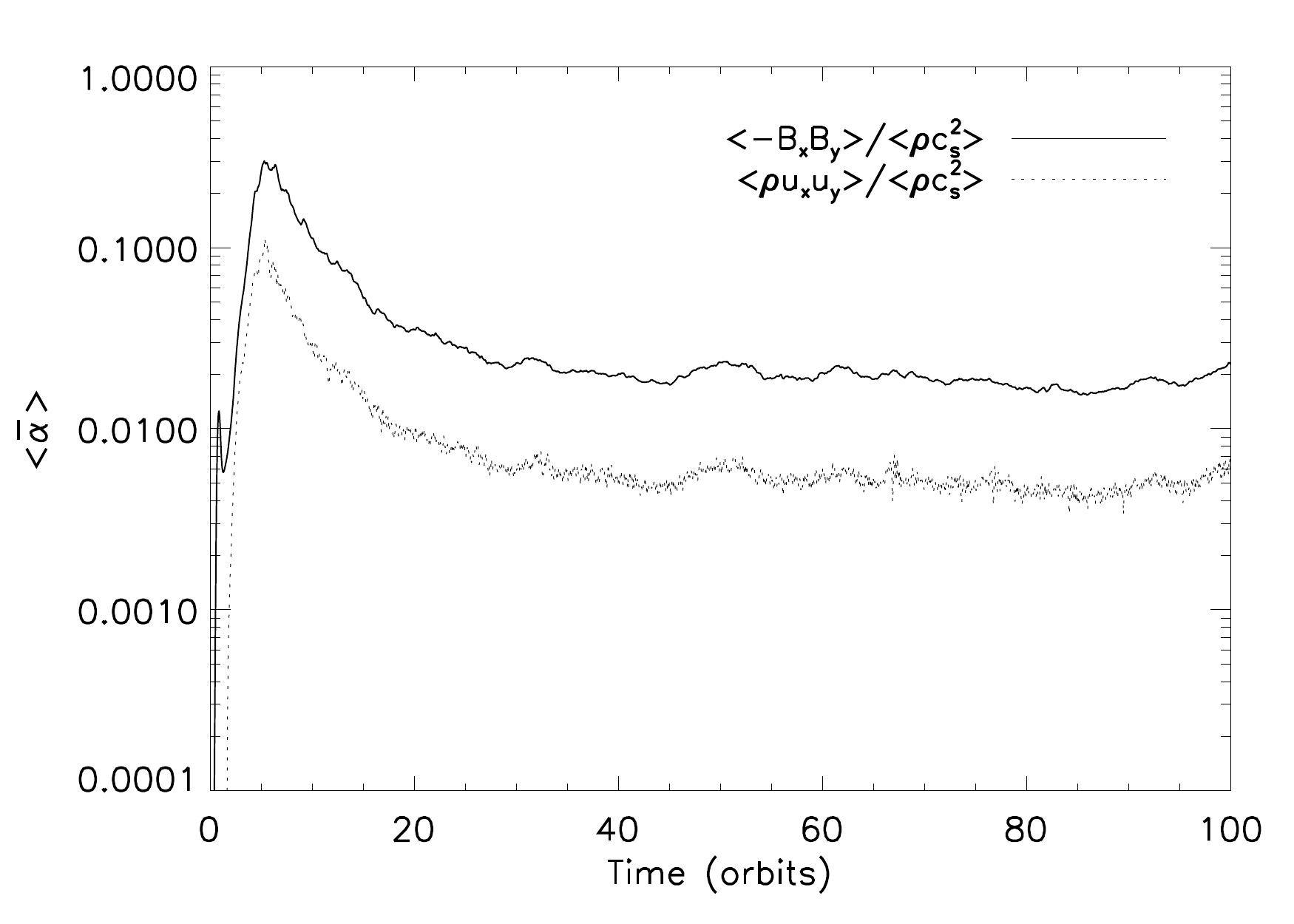}
	\caption{The upper diagram shows the time evolution of the magnetic energy density evaluated for the $B_x$ (solid line), $B_y$ (dotted line) and $B_z$ (dashed line) components. The bottom diagram shows the time evolution of the \citet{shakura_sunyaev_73} viscosity parameter $\left<\overline{\alpha} \right>$, where the solid line gives the contribution of the Maxwell stress tensor $\left<T_{xy}^{Max}\right>$, and the dotted line the contribution of the Reynolds stress tensor $\left<T_{xy}^{Rey}\right>$. In both diagrams, volume averages have been obtained from the entire computational domain.}
	\label{fig:21H_hstme}
\end{figure}

Figure \ref{fig:21H_hstme} shows the evolution of the \added{volume-averaged} magnetic energy density $\langle B^2\rangle/2$ (top diagram), and the \added{$\alpha$ parameter} (bottom diagram) over $100$ orbital periods. 
The top diagram indicates an amplification of the $B_{x}$ and $B_{z}$ components during the first $5$ orbital periods. Estimating the exponential growth time from  the linear perturbation solutions for the PRTI  obtained by \citet{kim_etal_97}, for the higher \added{altitudes} of the system ($3H-6H$), we find values ($\sim 1.5P-4.5P$) that are in agreement with our simulations, although  \citet{kim_etal_97} have neglected  differential rotation in their evaluation.  
After $5$ orbital periods, the total magnetic field decreases due to expansion and upward transport through the outflow boundaries in the vertical direction ``$z$''. A small dissipation due to turbulent magnetic reconnection probably also contributes to this decrease \added{(see more details in section \ref{sec:mag_rec})}.
At $t \sim 30P$, the  decrease of the magnetic field intensity and consequent increase of $\beta$ induces the growth of the MRI (from the azimuthal and vertical magnetic field components) that will dominate the system evolution for the rest of the simulation.

The  bottom diagram of Figure \ref{fig:21H_hstme} shows that  $\alpha$  is dominated by the Maxwell stress tensor and follows the same trend of the magnetic energy density, \added{as we expected}. \added{The diagram} also indicates that there is a high accretion regime in the first $10$ orbital periods when it achieves a maximum value around \added{$\left<\overline{\alpha} \right> \sim 0.3$} \citep[consistent with those expected from the observations, see ][]{king_etal_07}.

\subsubsection{Accretion disk and corona evolution}

\begin{figure}
	\includegraphics[scale=0.91]{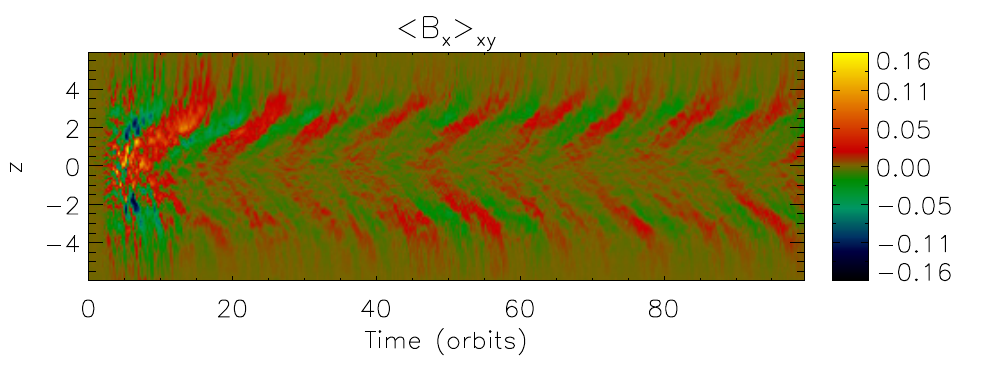}
	\includegraphics[scale=0.91]{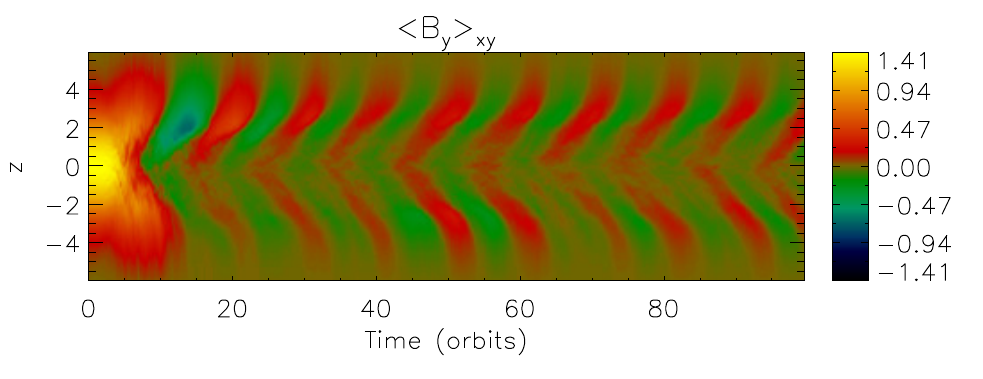}
    \includegraphics[scale=0.91]{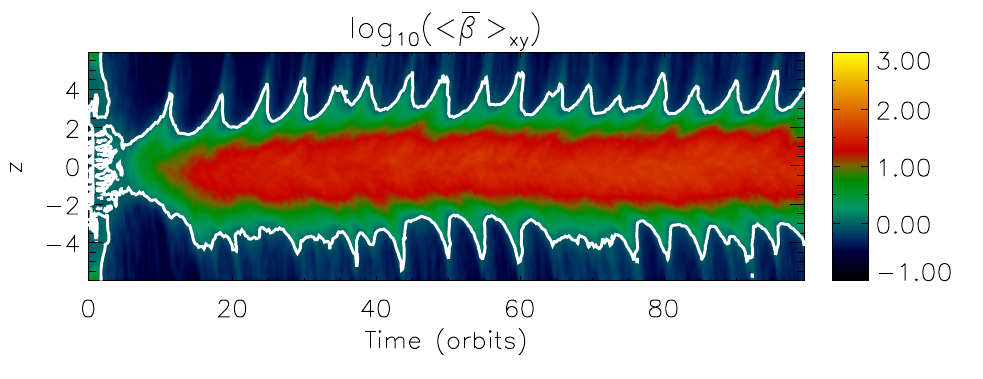}
	\caption{Time evolution and vertical profile of the horizontal averages of $B_x$, $B_y$, and $\beta$. The white line of the third diagram corresponds to  $\langle \overline{\beta} \rangle_{xy} =1$. It is possible to verify that after $10$ orbital periods, $B_x$ and $B_y$ present periodic variations with polarity inversions every $\sim 10$ orbital periods which are due to dynamo action triggered by the MRI.}
	\label{fig:21H_zpbm}
\end{figure}

\added{In order to evaluate the evolution of the accretion disk and the corona separately, we have obtained the horizontal averages of $B_x$ and $B_y$ magnetic field components\footnote{The evolution of the horizontal average of $B_{z}$ results a nearly zero value in different heights ($\langle B_{z} \rangle_{xy} \sim 10^{-7}$), though the volume average $\langle B_{z}^2 \rangle$ grows continuously with time during the PRTI regime (as indicate in Figure \ref{fig:21H_hstme})}, and $\beta$ parameter as a function of the height of the system.}
The first and second diagrams of the Figure \ref{fig:21H_zpbm} show a change of regimes between $t=5P$ and $t=10P$ of the radial and azimuthal components of the magnetic field. Besides, during the regime where the PRTI dominates, there is an increase of the radial component with a peak of intensity in the middle of the disk, followed by an inversion of polarity in the coronal regions between $|z|=2H$ and $4H$. In the higher \added{altitudes} of the corona, between $|z|=4H$ and $6H$, there are smaller polarity inversions.

After $t=10P$, the first and second diagrams of Figure \ref{fig:21H_zpbm} show that the radial and the azimuthal components of the magnetic field undergo polarity inversions on a time scale of approximately $10$ orbital periods. This is also a well-known result consistent with previous studies of MRI in weak field regimes ($\beta \gg 1$) and zero net flux \citep[e.g.,][]{brandenburg_etal_95, davis_etal_10, simon_etal_2011, simon_etal_2012, shi_etal_06}, which have produced similar butterfly diagrams to those we see in Figure \ref{fig:21H_zpbm}. This pattern suggests the action of an alpha-Omega dynamo process triggered  both by differential rotation (Omega effect) and the turbulent cyclonic motions driven by the PRTI and MRI (alpha effect). The first one stretches the radial and vertical magnetic field lines\footnote{Note that since the mean vertical $\langle B_z \rangle$ is null, it should not effectively participate in the amplification of the large-scale components, except possibly at the beginning of the evolution of the dynamo process in the PRTI phase.} in the azimuthal direction. The alpha effect allows for the amplification of the radial component ``$B_x$'' from the azimuthal component ``$B_y$'', through the electromotive forces in the azimuthal direction $(\boldsymbol{u} \times \boldsymbol{b})_y$, where $\boldsymbol{u}$ and $\boldsymbol{b}$ correspond to the fluctuations of velocity and magnetic field, respectively.
We also note that in Figure \ref{fig:21H_zpbm}, the polarity inversions of $B_x$ and $B_y$ components at the end of each half cycle (where the old field is replaced by a new one with opposite sign) are in phase opposition, as expected in a shear dynamo process \citep[for details, see, e.g.,][and references therein]{brandenburg_etal_95, guerrero_dgdp_07a, guerrero_dgdp_07b, guerrero_dgdp_08, guerrero_etal_09, guerrero_etal_16a, guerrero_etal_16b}. 

\added{Finally, the third diagram of Figure \ref{fig:21H_zpbm} shows the time evolution of $\langle \overline{\beta} \rangle_{xy}$. In the coronal region, the intensity of $\langle \overline{\beta} \rangle_{xy}$ decreases due to the growth of the magnetic field strength in this region induced by the PRTI. Thereafter, it saturates around an average value $\langle \overline{\beta} \rangle_{xy} \sim 0.8$ with a strong variability (between $\sim 0.5$ and $1.0$). Besides, even with the initial highly magnetized system ($\beta_{0}=1$), the disk evolves to a gas-pressure dominated regime with a horizontal-average value around $10$. Once the thermal pressure becomes larger than the magnetic pressure ($\beta>1$), the MRI is expected to settle in and soon dominate the dynamics of the disk. The MRI evolves out of the azimuthal and also of the vertical component of the magnetic field produced by the PRTI \citep[see also, e.g.,] []{balbus_hawley_92, hawley_etal_95, foglizzo_tagger_95, johansen_levin_08}.}

These results, although out of the main scope of this work, are compatible with the notion that the MRI starts to dominate the dynamics of the system when $\beta$ grows to values greater than the unit inside the disk. This inhibits the PRTI and at the same time gives rise to dynamo amplification of the azimuthal field and some amplification of the radial field which started to grow by dynamo process early in the PRTI regime.

\subsection{Comparison between models with different initial values of $\beta_{0}$}
\label{sec:comp_beta}

In this section, we show the evolution of the PRTI and MRI for different initial values of $\beta_{0}$ ($1$, $10$ and $100$) in a computational domain of size $12H \times 12H \times 12H$ and a resolution of $21H^{-1}$ (models R21b1, R21b10, and R21b100, respectively). The presence of a weak \added{vertical} magnetic field is essential for the development of the  MRI \citep[see, e.g.,][]{chandrasekhar_60, balbus_hawley_91, hawley_etal_95, miller_stone_2000, salvesen_etal_16a, salvesen_etal_16b}. In contrast, the PRTI evolves in regimes where the magnetic pressure is of the order of the thermal pressure \citep[$\beta_{0} \sim 1$; see, e.g.,][]{parker_66}. It is expected, therefore, that for systems with initially weak magnetic fields (or $\beta$ large) the PRTI should not dominate the initial evolution of the system. 

\begin{figure} 
	\includegraphics[scale=0.45]{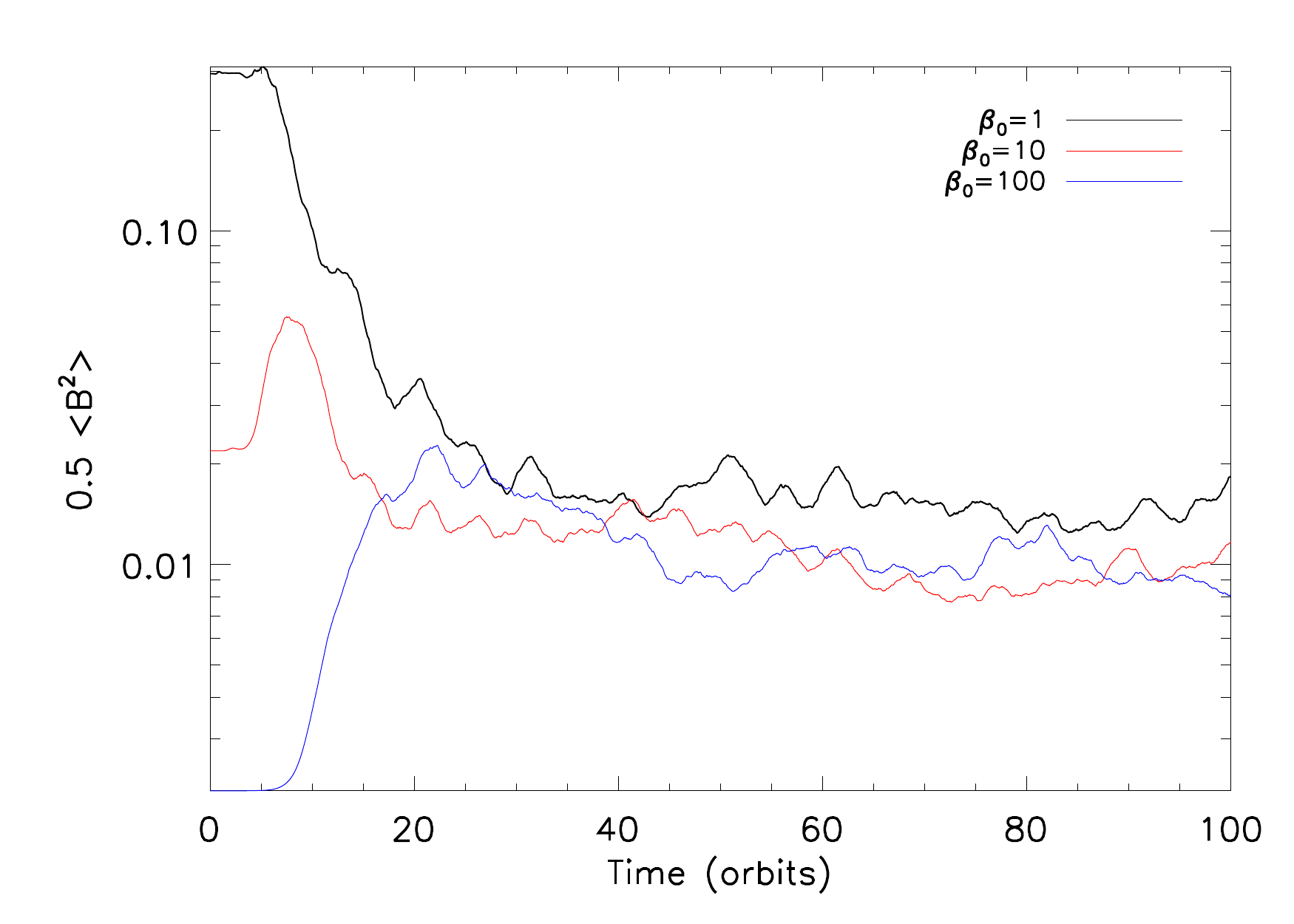}
    \includegraphics[scale=0.45]{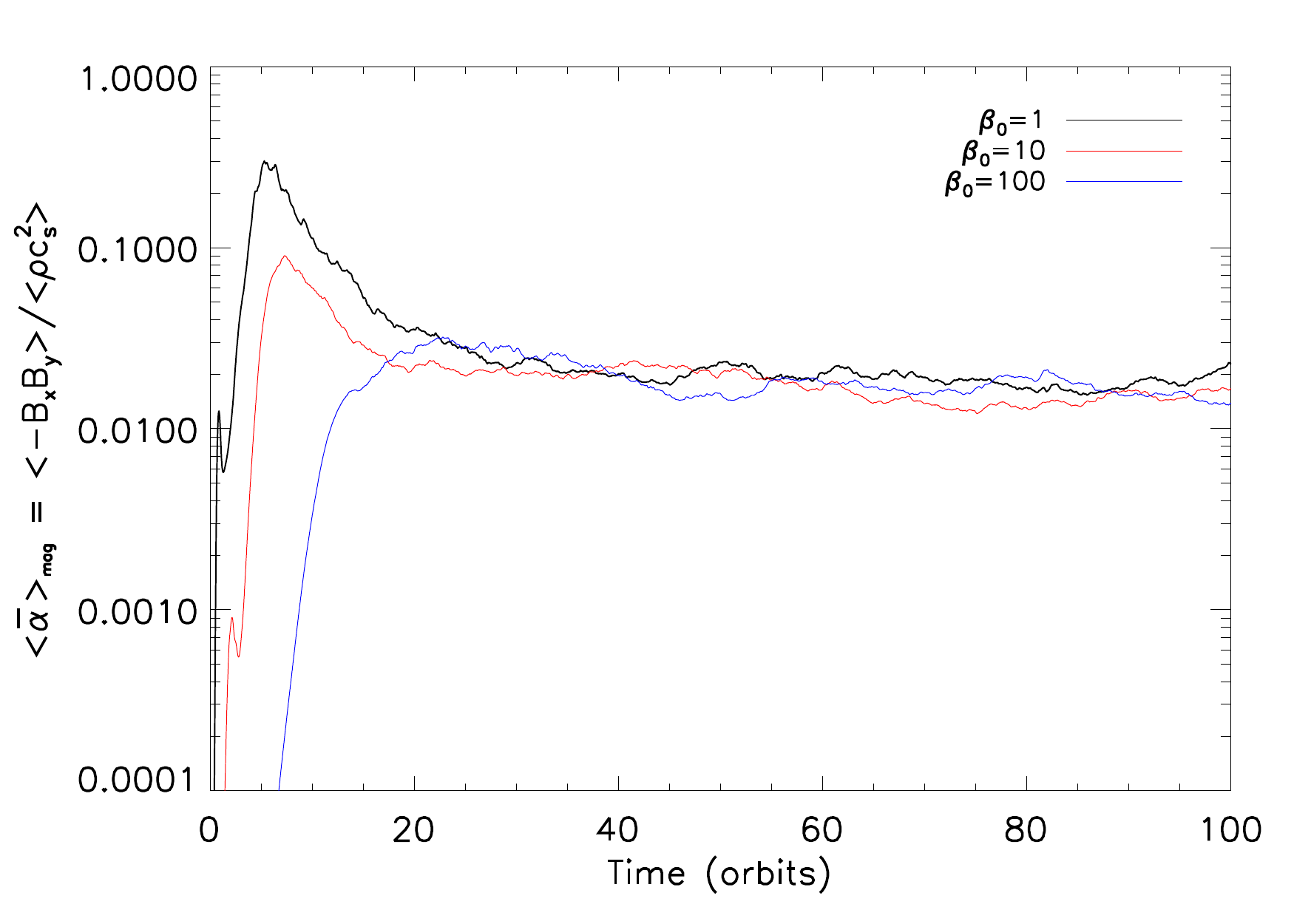}
    \caption{The first and second diagrams show the evolution of the volume-averaged total magnetic energy density ($\langle B^2 \rangle/2$) and $\langle\overline{\alpha}\rangle_{mag}=\langle-B_xB_y\rangle/\langle\rho c_s^2\rangle$, respectively. For all diagrams, the colors black, red, and blue correspond to the volume averages of these quantities obtained for different initial values of $\beta_{0}$. The black line also corresponds to the reference model (R21b1).}
\label{fig:comp_beta}
\end{figure}

The diagrams of Figure \ref{fig:comp_beta} show the time evolution of $\langle B^2 \rangle/2$ (top diagram) \added{and $\langle\overline{\alpha}\rangle_{mag}=\langle-B_xB_y\rangle/\langle\rho c_s^2\rangle$ (bottom diagram)}. The reference model R21b1 with $\beta_{0}=1$ discussed in the previous sections is also shown for comparison (black line). The top diagram shows clearly that the initial evolution of the magnetic field is different for models with distinct $\beta_{0}$. Nevertheless, after $20$ orbital periods all systems evolve in a similar way achieving a nearly steady state situation. For R21b10 model (with $\beta_{0}=10$), the initial amplification of the magnetic field and its subsequent decrease at the steady state phase indicates that the magnetic buoyancy still plays an important role in the transport of the magnetic field from the disk to outside the system through the vertical boundaries, while R21b100 model (with $\beta_{0}=100$) shows the standard evolution of the magnetic energy density driven by the MRI only, with an initial amplification until the saturation is achieved after $20$ orbital periods \citep[e.g.,][]{hawley_etal_95, miller_stone_2000}. 

\added{The $\alpha$ parameter} also shows convergence for all models after $20$ orbital periods (bottom diagram of Figure \ref{fig:comp_beta}). The initial peak value achieved during the growth phase of the PRTI seen in R21b1 model gets smaller for increasing $\beta_{0}$, as expected, and absent for $\beta_{0}=100$.

\begin{figure}
  \includegraphics[scale=0.91]{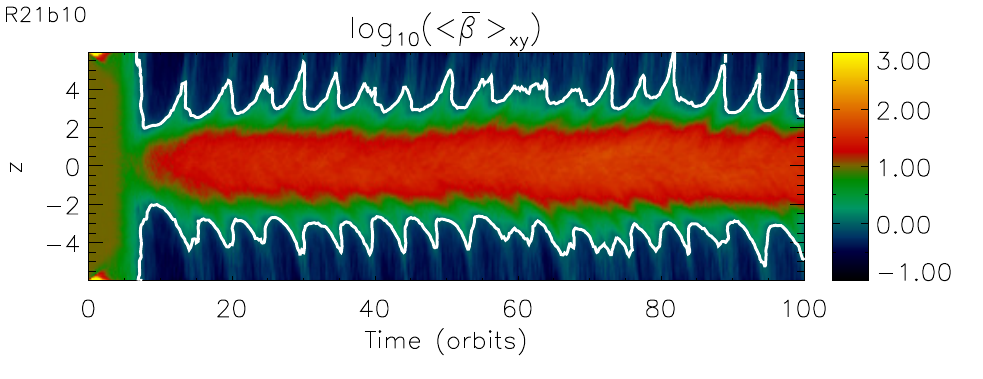}
  \includegraphics[scale=0.91]{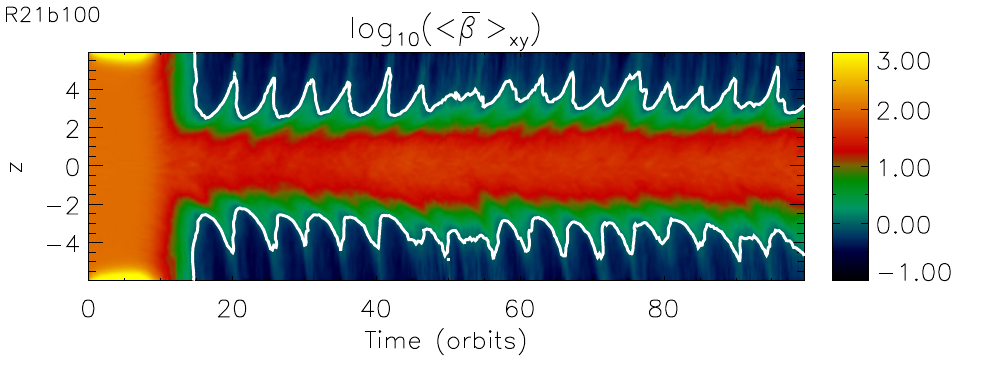}
  \caption{\added{Time evolution of $\langle \overline{\beta} \rangle_{xy}$ for the R21b10 (top diagram) and R21b100 (bottom diagram) models. The white line corresponds to the contour of $\langle \overline{\beta} \rangle_{xy} =1$.}}
  \label{fig:comp_beta01}
\end{figure}

The diagrams of Figure \ref{fig:comp_beta01} show the time evolution of $\langle \overline{\beta} \rangle_{xy}$ for the models discussed in this section. Consistent with the other diagrams, the initial growth of the magnetic field due to the PRTI in the models with  initial $\beta_{0} < 100$ leads to a decrease in $\beta$ and then, after $20$ orbital periods all the models converge to the same steady state values, both in the disk and corona, when the MRI sets in. At this stage, regardless of the initial $\beta_{0}$, the disk becomes a gas-pressure dominated system surrounded by a much more magnetized corona (with $\beta \simeq 1$), which is in agreement with \cite{miller_stone_2000, salvesen_etal_16b}. 
Finally, the differences found in all diagrams show that the PRTI has an important role just in an early ``transient'' phase.

\subsection{Numerical convergence}
\label{sec:num_conv}

As well known, the evolution of the MRI triggered by an initial azimuthal magnetic field is highly dependent on the resolution of the computational domain, since it is not possible to resolve all growing wavenumbers, specially the $k_z$ modes \citep[see][]{hawley_etal_95}. Besides, considering a zero net vertical flux in stratified shearing-box simulations, \citet{ryan_etal_17} have found that the $\alpha$ parameter is resolution dependent, which is in contrast to the results of, e.g., \cite{davis_etal_10}.
In this section, we will discuss the role of the resolution in the evolution of the PRTI and MRI and check whether the reference model (R21b1) converges numerically to a stationary state.

\begin{figure} 
  \includegraphics[scale=0.45]{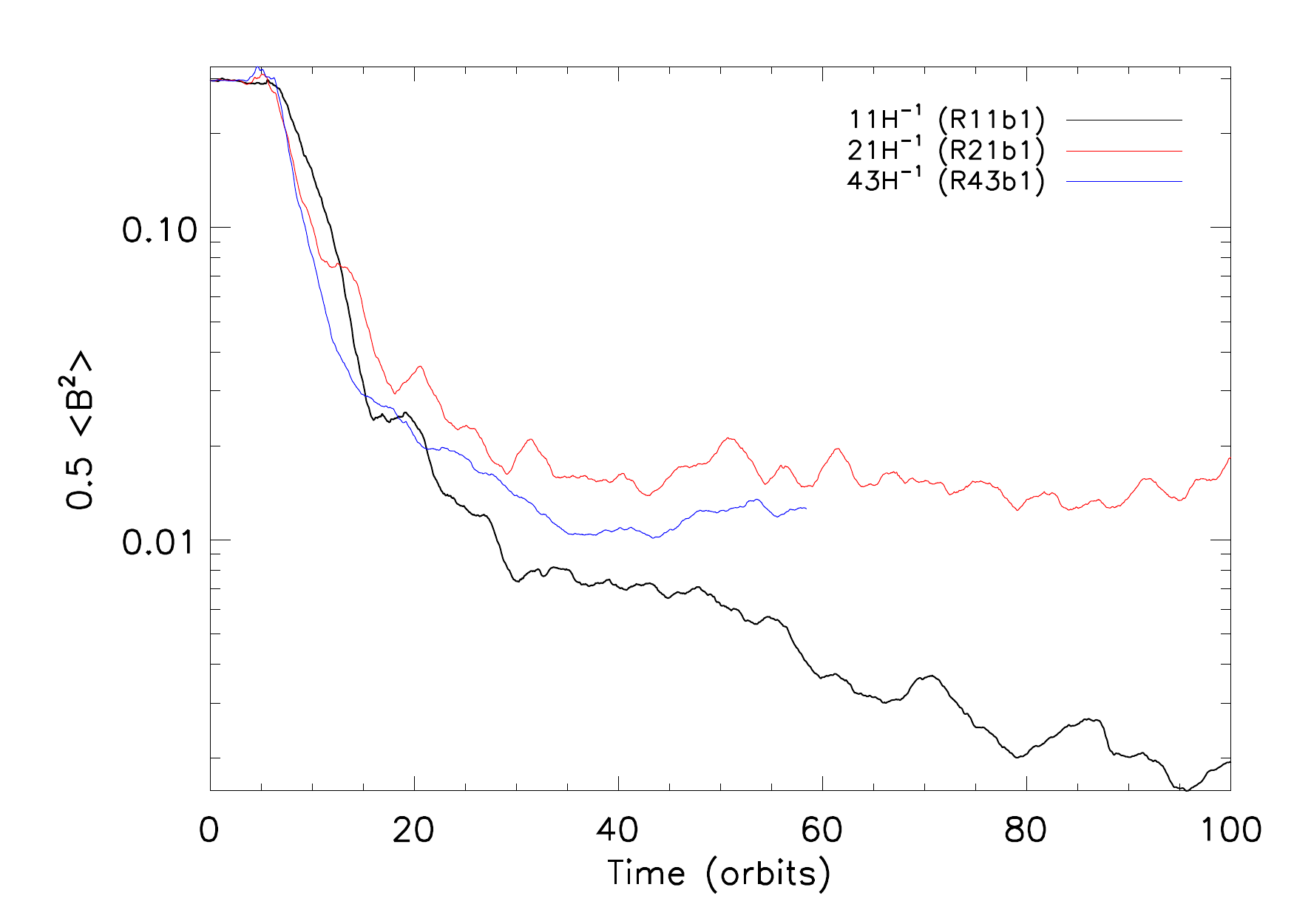}
  \includegraphics[scale=0.45]{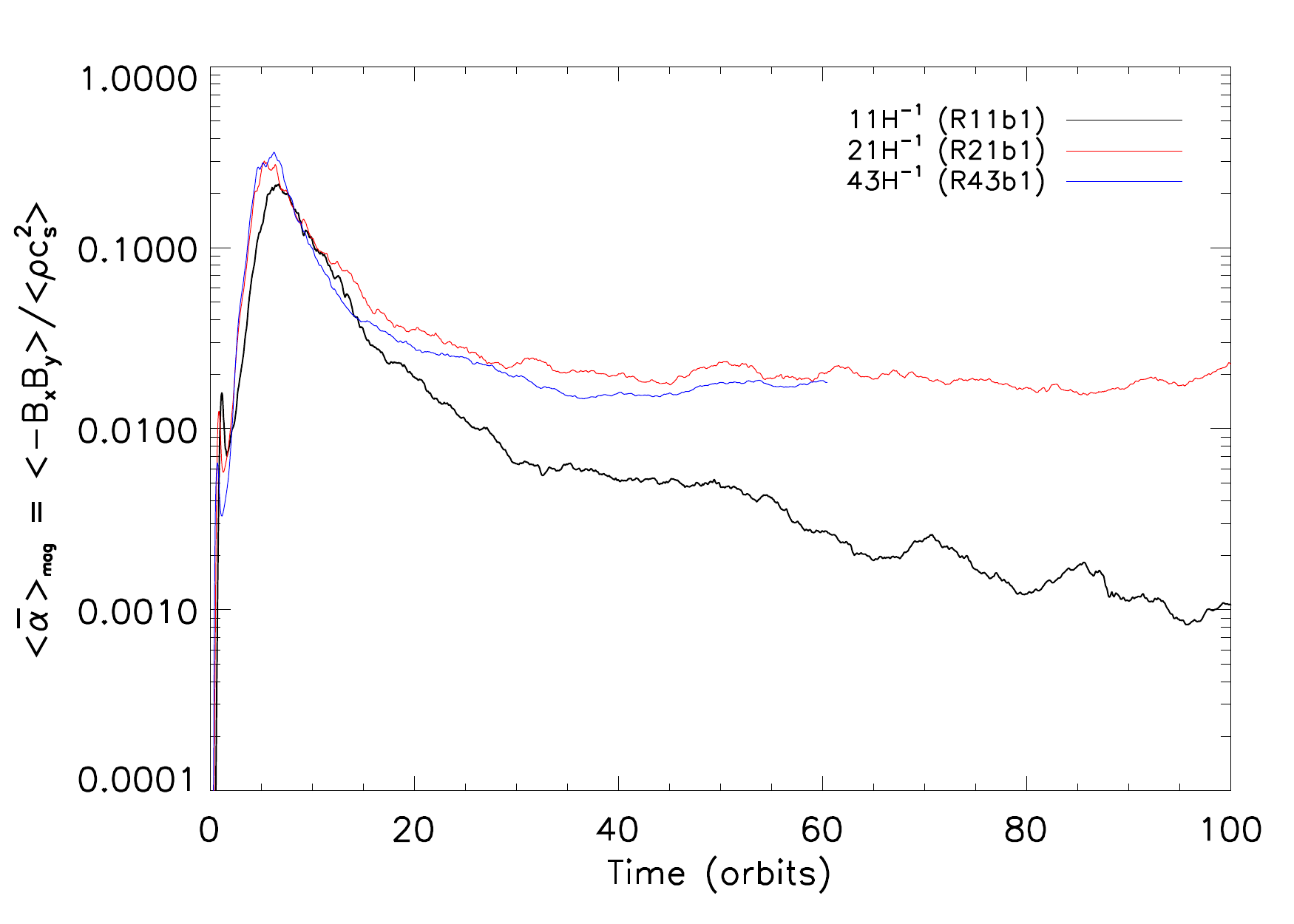}
\caption{Time evolution of the magnetic energy density (top diagram) and \added{ $\langle\overline{\alpha}\rangle_{mag}$} (bottom diagram) for the resolutions of $11H^{-1}$ (black line), $21H^{-1}$ (red line), and $43H^{-1}$ (blue line).}
\label{fig:comp_11H21H43H}
\end{figure}

 Figure \ref{fig:comp_11H21H43H} shows the time evolution of $\langle B^2 \rangle/2$ (top diagram) and $\langle\overline{\alpha}\rangle_{mag}$ (bottom diagram) for the resolutions of $11H^{-1}$ (R11b1 model, black line), $21H^{-1}$ (R21b1 model, red line), and $43H^{-1}$ (R43b1 model, blue line). The high-resolution simulation (R43b1) is numerically costly and we have evolved over a short time equivalent to $\sim 58$ orbital periods. During the first $5$ orbital periods (where the PRTI is dominant) until $t=20P$, all the resolutions show the same behavior for the magnetic energy density, indicating that the PRTI operates appropriately even in our lower-resolution simulation (R11b1). It is known that the growth rate of the PRTI is favored by large wavenumbers \citep[see, e.g.,][]{parker_66, parker_67, kim_etal_97}, so we might expect that the resolution of $11H^{-1}$ would not affect significantly the initial evolution of such instability. Nevertheless, the maximum value of $\langle\overline{\alpha}\rangle_{mag}$ for this model  is slightly lower than for the intermediate and high-resolution simulations (R21b1 and R43b1 models, respectively), indicating that resolutions lower than $11H^{-1}$ could affect the initial evolution of the system.

After $20$ orbital periods, when the MRI dominates, the differences in the evolution between the low and intermediate resolution models is significant. 
The low-resolution model continues to decrease the magnetic field without achieving saturation. Such decrease indicates that the MRI in this case is unable to regenerate the magnetic field at the same rate that magnetic flux is transported throughout the vertical boundaries.
As mentioned, such behavior is due to sensitivity of the MRI to the resolution when triggered by an initial azimuthal magnetic field \citep[see,][]{hawley_etal_95}. 
The comparison between the intermediate and high-resolution models indicate a good agreement between them until the evolved time of the high-resolution model R43b1 (around $58$ orbits), though it is not possible to determine yet whether it will achieve the same saturation as in  R21b1 model.

\section{The search for magnetic reconnection}
\label{sec:mag_rec}

The PRTI is fundamental in the formation of magnetic loops in the coronal region of the accretion disk. During this process, the encounter and squeezing between loops may lead to magnetic reconnection. Indeed, magnetic reconnection is expected to occur whenever two magnetic fluxes of opposite polarity approach each other in the presence of finite magnetic resistivity. The ubiquitous microscopic Ohmic resistivity is enough to allow for magnetic reconnection, though in this case the rate at which the lines reconnect is expected to be very slow according to the Sweet-Parker mechanism \citep{sweet_58,parker_57}. In the present analysis, we are dealing with ideal MHD simulations with no explicit resistivity. This in principle would prevent us from detecting  magnetic reconnection. Nevertheless, in numerical MHD simulations magnetic reconnection can be excited because of the presence of numerical resistivity \added{that can mimic the Ohmic resistivity of real plasmas}. This in turn, could make one to speculate that the identification of potential sites of magnetic reconnection in ideal MHD simulations would be essentially a numerical artifact. Nevertheless, according to the Lazarian-Vishniac reconnection model \citep{lazarian_vishiniac_99, kowal_etal_09,santosLima_etal_10,eyink_etal_11}, the presence of turbulence in real MHD flows is able to speed up the reconnection to rates nearly as large as the local Alfv\'{e}n speed \added{and independent of the Ohmic resistivity (which in numerical simulations is replaced by the numerical resistivity)}. This occurs because of the turbulent wandering of the magnetic field lines that allow them to encounter each other in several patches simultaneously making reconnection very fast. Even in sub-Alfv\'{e}nic flows, where the magnetic fields are very strong, this fast reconnection process may be very efficient. \added{This model  and the independence of the fast recconection rate with the Ohmic (or numerical) resistivity (see eq.\ref{eq:vrec_turb}) has been extensively and successfully tested numerically in several studies involving ideal MHD simulations of turbulent flows \citep[see, e.g., the references above and the reviews of][]{lazarian12, lazarian15, dgdp_etal_15}.} 

We have seen here that the PRTI and the MRI instabilities are able to trigger turbulence in the flow and also the close encounter of magnetic field lines with opposite polarity in several regions and specially at the coronal loops. We will see below that these regions are loci of large increase of the current density in very narrow regions and are, therefore, potential sites of magnetic discontinuities or current sheets. As stressed above, the numerical resistivity here simply mimics the effects of the finite Ohmic resistivity, but the important physical real mechanism that may allow for the fast reconnection is the process just described due to the turbulence.

\subsection{Method}
\label{sec:met_vrec}

In order to identify magnetic reconnection sites in our disk/corona system, we have searched for current density peaks ($\boldsymbol{J} = \nabla \times \boldsymbol{B}$) and then evaluated the local magnetic reconnection rate ($V_{rec}$) in the surroundings of each peak.
To this aim, we have adapted the algorithm of \citet{zhdankin_etal_13} and extended it to a 3D analysis. The algorithm is well described in \citet{zhdankin_etal_13} and we will summarize the most important steps in this section. First, we have selected a sample of cells with a current density value greater than $j_{5}= 5\langle |\boldsymbol{J}| \rangle$, where $\langle |\boldsymbol{J}| \rangle$ is the volume average of the total current density taken over the disk and the corona separately. From this first sample, we have selected those cells that have local maxima ($|\boldsymbol{J}|_{ijk}^{max}$, where $ijk$ corresponds to the index position of each cell) within a surrounding cubic subarray of the data (with a size of $7 \times 7 \times 7$ cells) and then we  checked whether they are located between  magnetic field lines of opposity polarity (in each component separately). In the present work, we are not interested to identify null points since the magnetic reconnection topology is complex in a 3D regime \citep[see, e.g.,][]{yamada_etal_10}. Besides, this step is quite different from the original algorithm because \citet{zhdankin_etal_13} have used an X-point model \citep[see][]{petschek_64} to represent the magnetic reconnection sites and then they identified such events as saddle points in the magnetic flux function.

\begin{figure} 
	\includegraphics[scale=0.31]{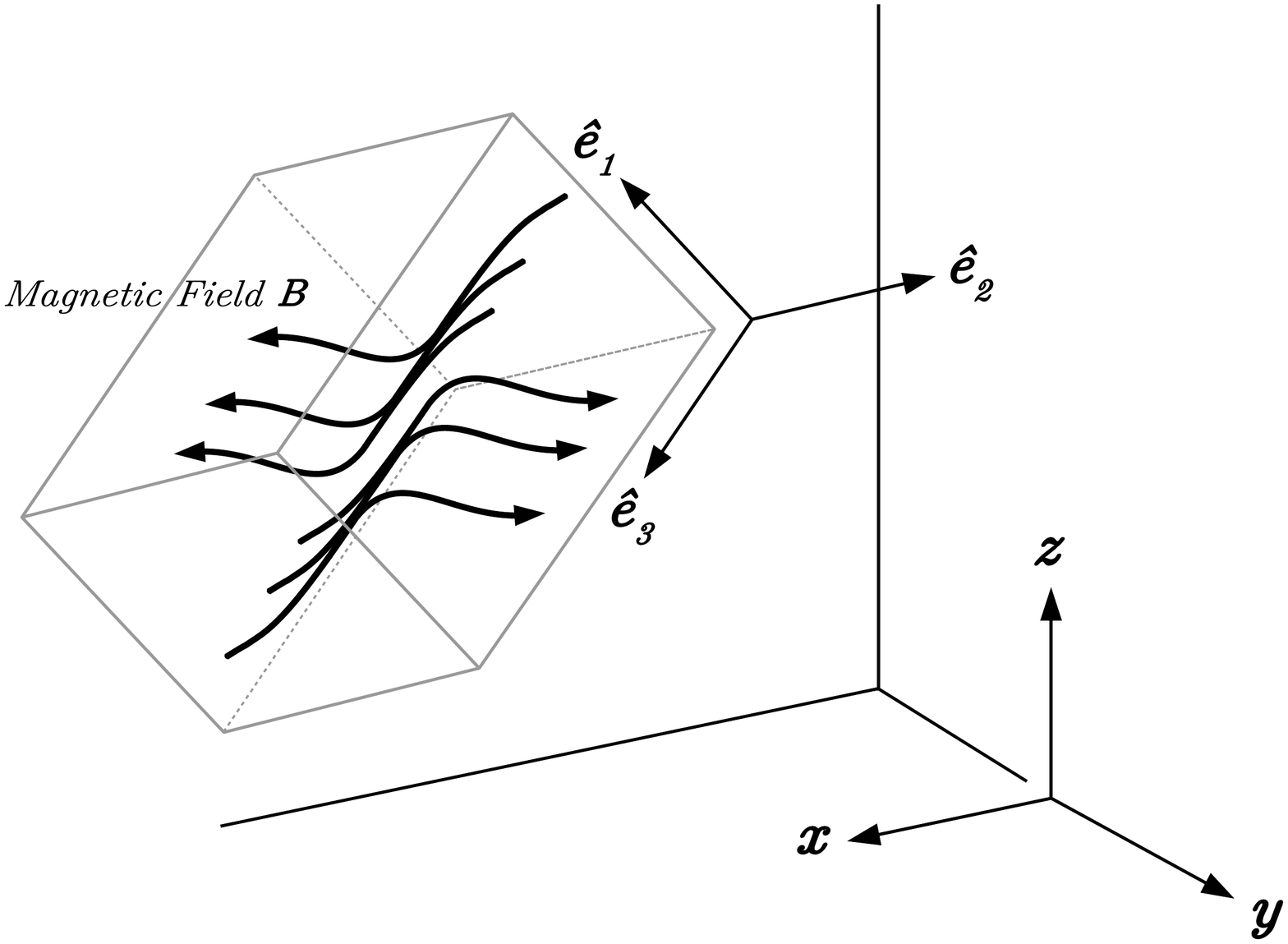}
	\includegraphics[scale=0.31]{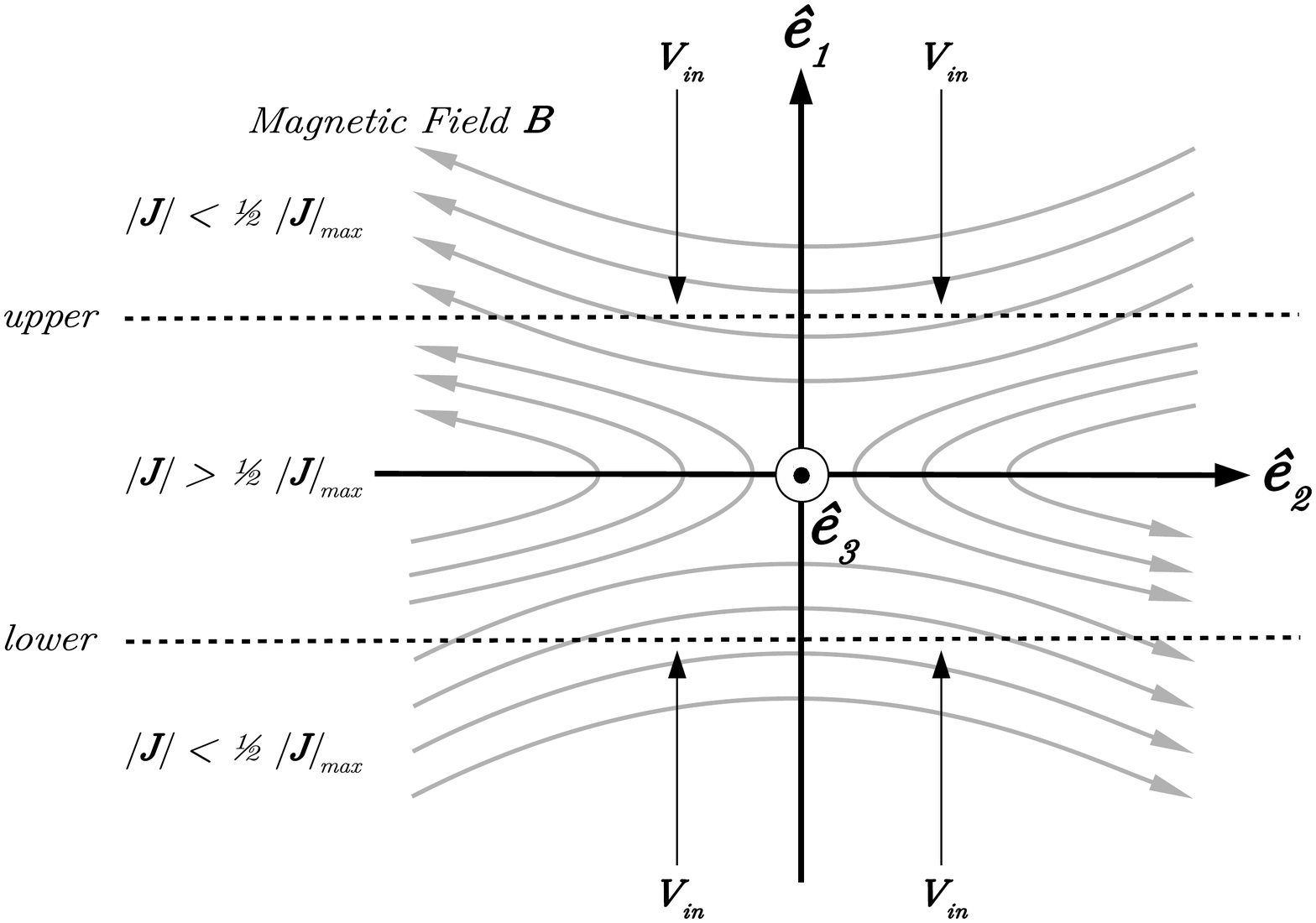}
	\caption{The upper sketch shows a new coordinate system centered in a local reconnection site. We have identified the orthonormal vector $\boldsymbol{\hat{e}_1}$ as the direction of the fastest decrease of $|\boldsymbol{J}|_{ijk}^{max}$. We have assumed this direction as the thickness of the magnetic reconnection site. The bottom sketch shows the details of the reconnection configuration, where the edges have been defined as the cells of $|\boldsymbol{J}| <1/2 |\boldsymbol{J}|_{ijk}^{max}$.}
	\label{fig:eigenvectors}
\end{figure}

We have assumed that the cells of the last subsample are possible sites of reconnection. However, the topology of these sites is complex (as mentioned above) and they are not necessarily aligned with one of the axes of the Cartesian coordinate system. So, we adopted  a new coordinate system centered in the local reconnection site obtained from the eigenvalues and eigenvectors of the current density 3D Hessian matrix ($H_{ijk}$) for each cell of the last subsample \citep[see][]{zhdankin_etal_13}:

\begin{equation}
H_{ijk}=\left[ {\begin{array}{ccc}
	
	       \partial_{xx}|\boldsymbol{J}|_{ijk} & \partial_{xy}|\boldsymbol{J}|_{ijk} & \partial_{xz}|\boldsymbol{J}|_{ijk} \\
	       \partial_{yx}|\boldsymbol{J}|_{ijk} & \partial_{yy}|\boldsymbol{J}|_{ijk} & \partial_{yz}|\boldsymbol{J}|_{ijk} \\
	       \partial_{zx}|\boldsymbol{J}|_{ijk} & \partial_{zy}|\boldsymbol{J}|_{ijk} & \partial_{zz}|\boldsymbol{J}|_{ijk} \\
	 
	      \end{array}} \right] ,
\end{equation}
where $H_{ijk}$ corresponds to the second-order partial derivatives of the current density magnitude. The highest eigenvalue indicates that the associated eigenvector corresponds to the direction of the fastest decrease (or the highest variance) of $|\boldsymbol{J}|$. We have assumed this direction as the thickness of the magnetic reconnection site. The eigenvectors of $H_{ijk}$ provide the three orthonormal vectors of the new coordinate system centered in the local reconnection site (see the upper sketch of Figure \ref{fig:eigenvectors}).   

\begin{figure}
    \centering
	\includegraphics[scale=0.41]{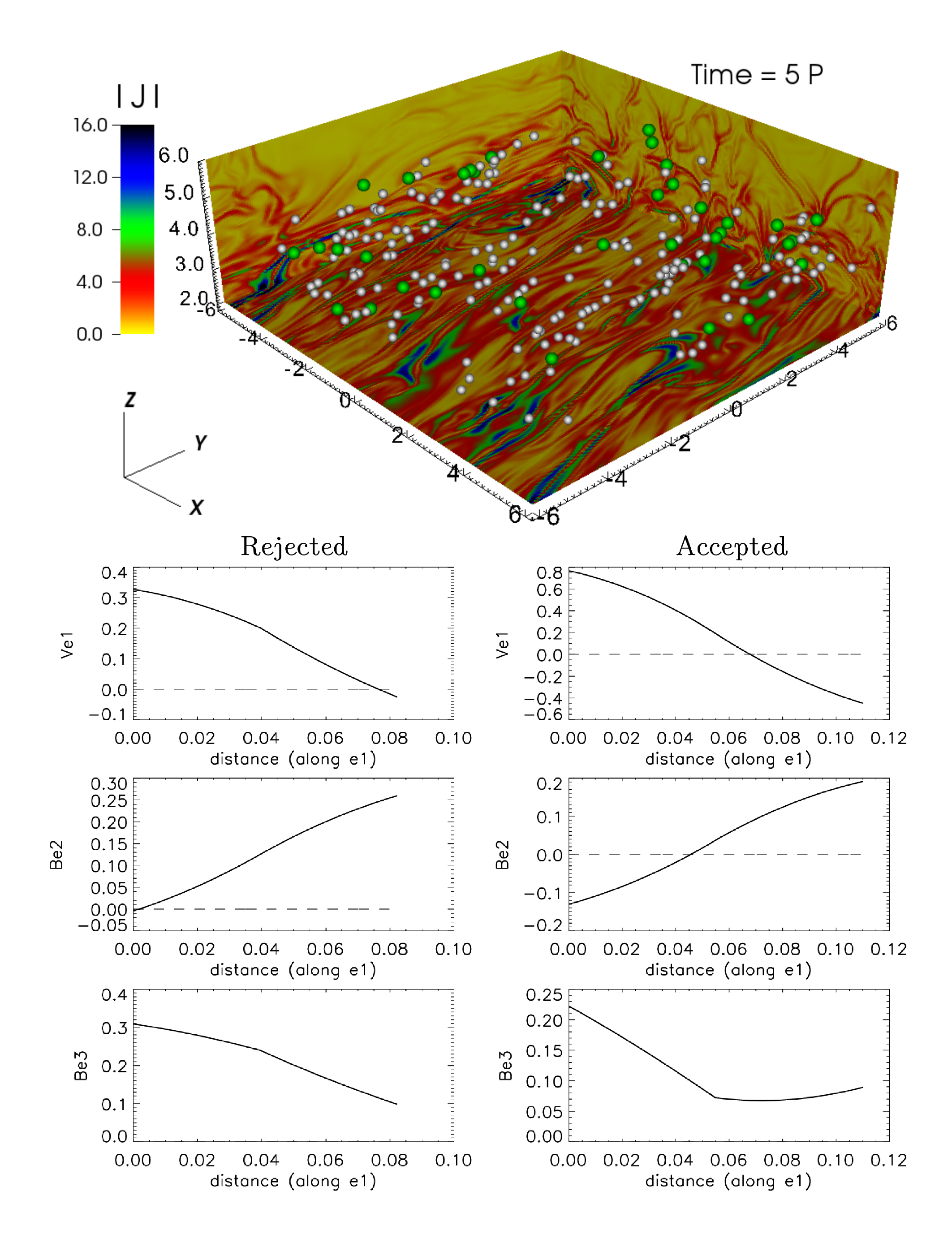}
	\caption{Top: diagram of the coronal region above the disk (upper corona, $3H<z<6H$) at $t=5P$ for R21b1 model. The colors in the background correspond to the current density magnitude. The white points correspond to the local maxima identified by the algorithm and the green points correspond to the magnetic reconnection sites with the most symmetric magnetic and velocity field profiles. Bottom: the diagrams show examples of rejected and accepted profiles of $V_{\boldsymbol{\hat{e}_1}}$, $B_{\boldsymbol{\hat{e}_2}}$ and $B_{\boldsymbol{\hat{e}_3}}$ interpolated along the $\hat{e}_1$ axis. Those evidencing substantial symmetry in velocity and magnetic field in the separatrix have been accepted.}
	\label{fig:21H_recpos}
\end{figure}

The local magnetic reconnection rate has been evaluated in a similar way as in \citet{kowal_etal_09}, where we averaged the inflow velocity ($V_{in}=V_{\boldsymbol{\hat{e}_1}}$, the projection onto the $\boldsymbol{\hat{e}_1}$ direction) divided by the Alfv\'{e}n speed at the edges of the reconnection site (see the bottom sketch of Figure \ref{fig:eigenvectors}):

\begin{equation}
\left\langle {V_{in}\over V_{A}} \right\rangle = {1\over 2} \left({V_{\boldsymbol{\hat{e}_1}} \over V_{A}}\bigg\vert_{lower} -
{V_{\boldsymbol{\hat{e}_1}} \over V_{A}}\bigg\vert_{upper} \right) ,
\label{eq:vrec}
\end{equation}
where the Alfv\'{e}n speed is given by:

\begin{equation}
V_{A} = \sqrt{B_{\boldsymbol{\hat{e}_1}}^2 + B_{\boldsymbol{\hat{e}_2}}^2 + B_{\boldsymbol{\hat{e}_3}}^2 \over \rho},
\end{equation}
and $B_{\boldsymbol{\hat{e}_1}}$, $B_{\boldsymbol{\hat{e}_2}}$, and $B_{\boldsymbol{\hat{e}_3}}$ correspond to the projection of the magnetic field onto the three eigenvectors of the Hessian matrix. As in \citet{zhdankin_etal_13}, we have identified the edges and the cells  belonging to a given local magnetic reconnection site considering only those with current density $|J|$ greater than half of the maximum local value  given by $|\boldsymbol{J}|_{ijk}^{max}$
($|J| > 1/2 |\boldsymbol{J}|_{ijk}^{max}$, see the bottom sketch of Figure \ref{fig:eigenvectors}). We have constrained the subsample to obtain $V_{rec}$ for the most symmetric profiles, since it is expected a symmetry of both the magnetic and velocity field profiles around the magnetic reconnection site (as seen in Figure \ref{fig:eigenvectors}). As an example, Figure \ref{fig:21H_recpos} shows the spatial distribution of the local maxima identified by the algorithm, in the coronal region (upper corona, $3H<z<6H$) at $t=5P$ for the R21b1 model. The white points correspond to the total subsample and the green points to the constrained subsample, as described above. The diagrams of Figure \ref{fig:21H_recpos} show examples of rejected and accepted profiles of $V_{\boldsymbol{\hat{e}_1}}$, $B_{\boldsymbol{\hat{e}_2}}$ and $B_{\boldsymbol{\hat{e}_3}}$ interpolated along the $\hat{e}_1$ axis. This constraint reduces considerably the subsample size, but also reduces \added{the standard deviation of $V_{rec}$ (see a detailed discussion in sections \ref{sec:vrec} and \ref{sec:comp_vrec})}.

\begin{figure}
    \includegraphics[scale=0.85]{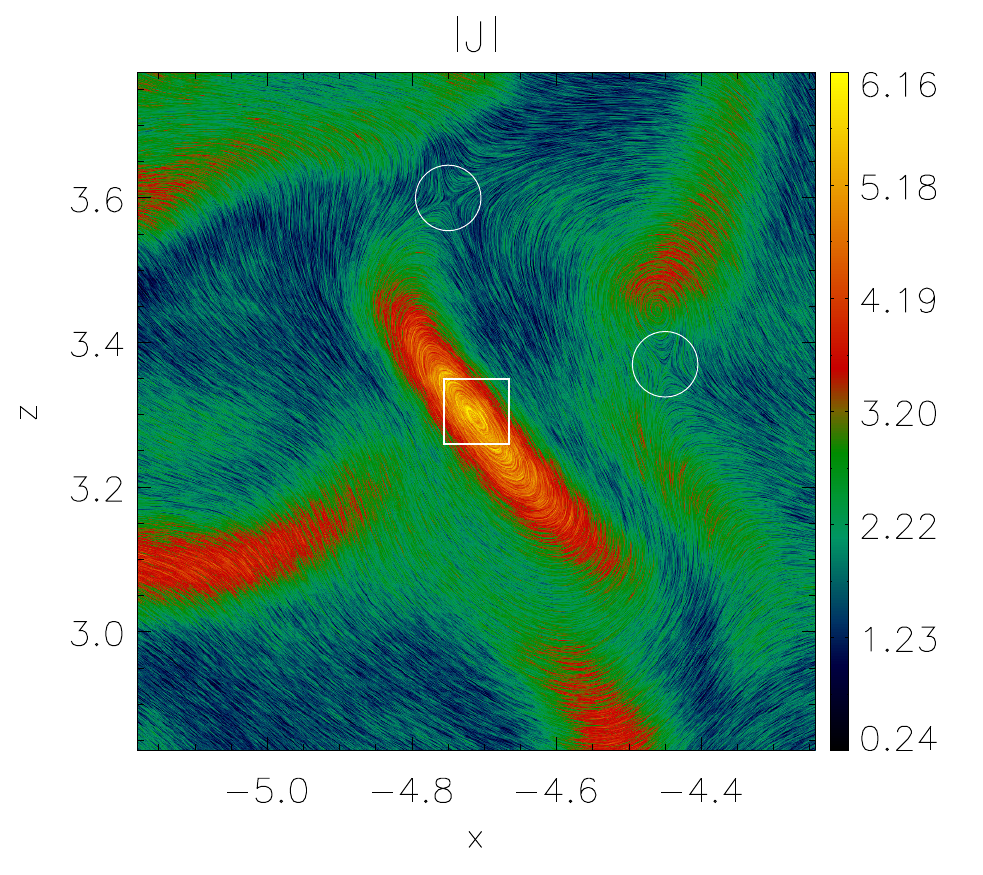}
    \includegraphics[scale=0.85]{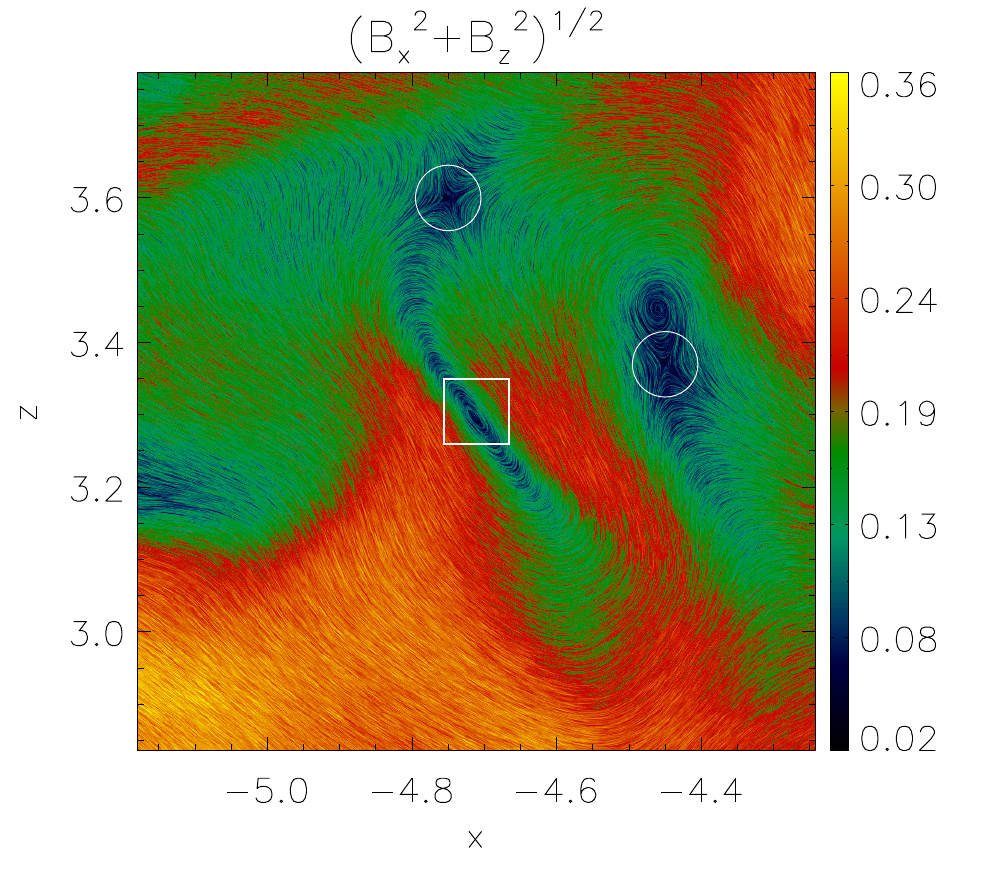}
    \caption{\added{The diagrams show the magnetic field lines in a region of the domain produced by a Line Integral Convolution (LIC) method combined with the maps of the total current density ($|\boldsymbol{J}|$, top diagram) and the magnetic field magnitude ($B_x$ and $B_z$ components, bottom diagram). The diagrams have been obtained interpolating the data in the surrounds of the reconnection region (white square) identified by the algorithm. The white circles correspond to other possible reconnected sites.}}
    \label{fig:lic_rec}
\end{figure}

\added{Finally, despite the challenge to represent the real topology of the reconnection site in our 3D-MHD simulations, as mentioned above, Figure \ref{fig:lic_rec} shows an example of one of the accepted regions using the Line Integral Convolution (LIC) method combined with 2D maps of the current density (top diagram) and the magnetic field intensity (bottom diagram) at $t=5P$. The diagrams have been obtained interpolating the data in the surrounds of the reconnection site and correspond to areas of $21$ cells$^2$. In this example, the $\hat{e}_3$ axis of the reconnection region (see Figure \ref{fig:eigenvectors}) is approximately aligned to the $y$-axis of the Cartesian coordinate system, allowing us to visualize the topology in the $xz$ plane. The magnetic field magnitude was evaluated using the $B_x$ and $B_z$ components. The bottom diagram of Figure \ref{fig:lic_rec} shows at least three possible reconnection events (white square and circles) at regions with   convergence of the magnetic lines where the magnetic intensity decreases. However, the algorithm identified only one of these regions (white square), that corresponds to the local maxima of $|\boldsymbol{J}|$ (see the top diagram of Figure \ref{fig:lic_rec}). It is out of the scope of this work to provide a detailed analysis of all  possible reconnection sites, but we can speculate that the other two regions (circles)  correspond to already reconnected sites where substantial magnetic energy has already dissipated.}

\subsection{Magnetic field configuration}

\begin{figure}
\centering
\includegraphics[scale=0.5]{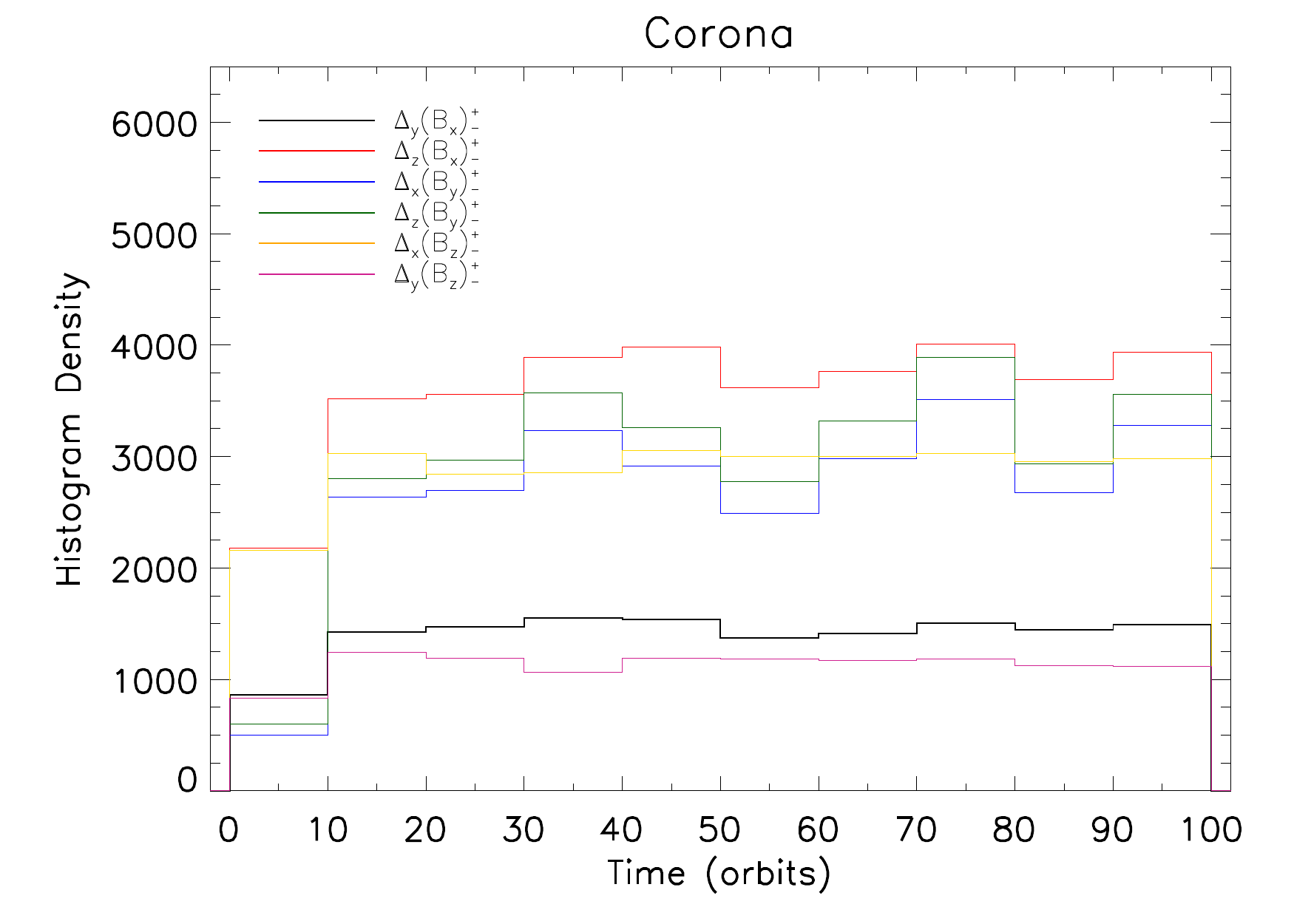}
\includegraphics[scale=0.5]{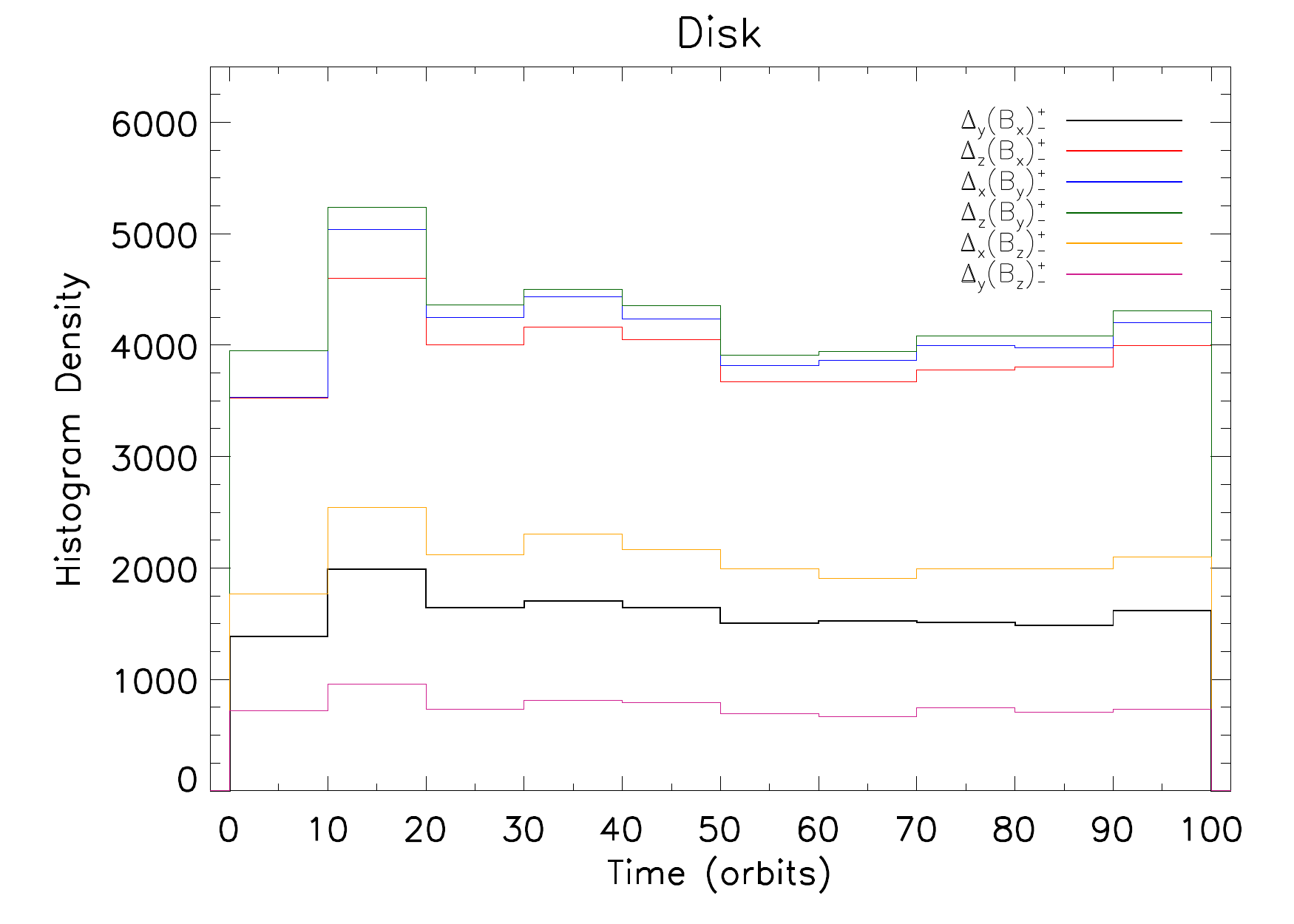}
\caption{Distributions of the sign's flip of the magnetic field components ``$i$'' in the direction ``$j$'' ($\Delta_{j}(B_{i})^{+}_{-}$). The counts have been organized in bins of $10$ orbital periods. The upper and lower histograms correspond to the distributions taken in the coronal region and disk, respectively. }
\label{fig:21H_histdb}
\end{figure}

As mentioned above, the algorithm checks whether each sample of cells is located between magnetic field lines of opposite polarity. We have also analyzed the distribution in time of the sign's flips of the magnetic field \added{ in each direction separately (``$x$'', ``$y$'', and ``$z$'') at spatial scales corresponding to the size of a cell.}
These distributions can help to identify what magnetic field components are reconnecting inside the system as a function of time, e.g., during the PRTI and MRI phases.
As an example, Figure \ref{fig:21H_histdb} shows such distributions (with bins of $10$ orbital period) for the model R21b1, where $\Delta_{j}(B_{i})^{+}_{-}$ corresponds to the number of sign's flip of the magnetic field component ``$i$'' in the direction ``$j$''.
The upper and lower diagrams correspond to the distribution taken in the coronal region and disk, respectively. Initially, between $0$ and $10$ orbital periods at the coronal region, the number of sign's flips is dominated mainly by the $B_{x}$ component along the vertical direction ``$z$'' ($\Delta_{z}(B_{x})^{+}_{-}$) and the $B_{z}$ component along the radial direction ``$x$'' ($\Delta_{x}(B_{z})^{+}_{-}$). This behavior indicates that reconnections of the azimuthal field are not relevant during this initial phase of the PRTI at the corona, showing that the field lines initially twist, producing flux tubes  in the azimuthal direction \citep[see also][]{simon_etal_2012,hirose_etal_96} and allowing for reconnection to occur in the ``$x$'' and ``$z$'' directions mainly. This behavior can be clearly seen in the diagrams of Figure \ref{fig:21H_ic}. 
On the other hand, after $t=10P$, the number of sign's flip of the $B_{y}$ component in the ``$x$'' and ``z'' directions ($\Delta_{x}(B_{y})^{+}_{-}$ and $\Delta_{z}(B_{y})^{+}_{-}$, respectively) increases  playing  an important role in the reconnection too, which is compatible with the inversion of polarity of the azimuthal magnetic field during the MRI phase (as seen in the middle diagram of Figure \ref{fig:21H_zpbm}). The sign's flips in the azimuthal direction ($\Delta_{y}(B_{x})^{+}_{-}$ and $\Delta_{y}(B_{z})^{+}_{-}$) are less relevant during all the evolution of the simulation, as expected, since the shear and stretching prevent local encounters of magnetic field lines of opposite polarity.

Similar behavior can be found in the disk region, except for the $B_{z}$ component in the ``$x$'' direction ($\Delta_{x}(B_{z})^{+}_{-}$), for which the flips are less relevant than in the coronal region. During all the simulation, the sign's flips are dominated by $\Delta_{z}(B_{y})^{+}_{-}$, $\Delta_{x}(B_{y})^{+}_{-}$ and $\Delta_{z}(B_{x})^{+}_{-}$.
This is not a surprise since the $B_{z}$ component is produced mainly by loops at the higher \added{altitudes} above and below the disk during the exponential growth of the PRTI (as described in section \ref{sec:num_results}).

\subsection{Magnetic reconnection rate $V_{rec}$}
\label{sec:vrec}

\begin{figure} 
\centering
\includegraphics[scale=0.5]{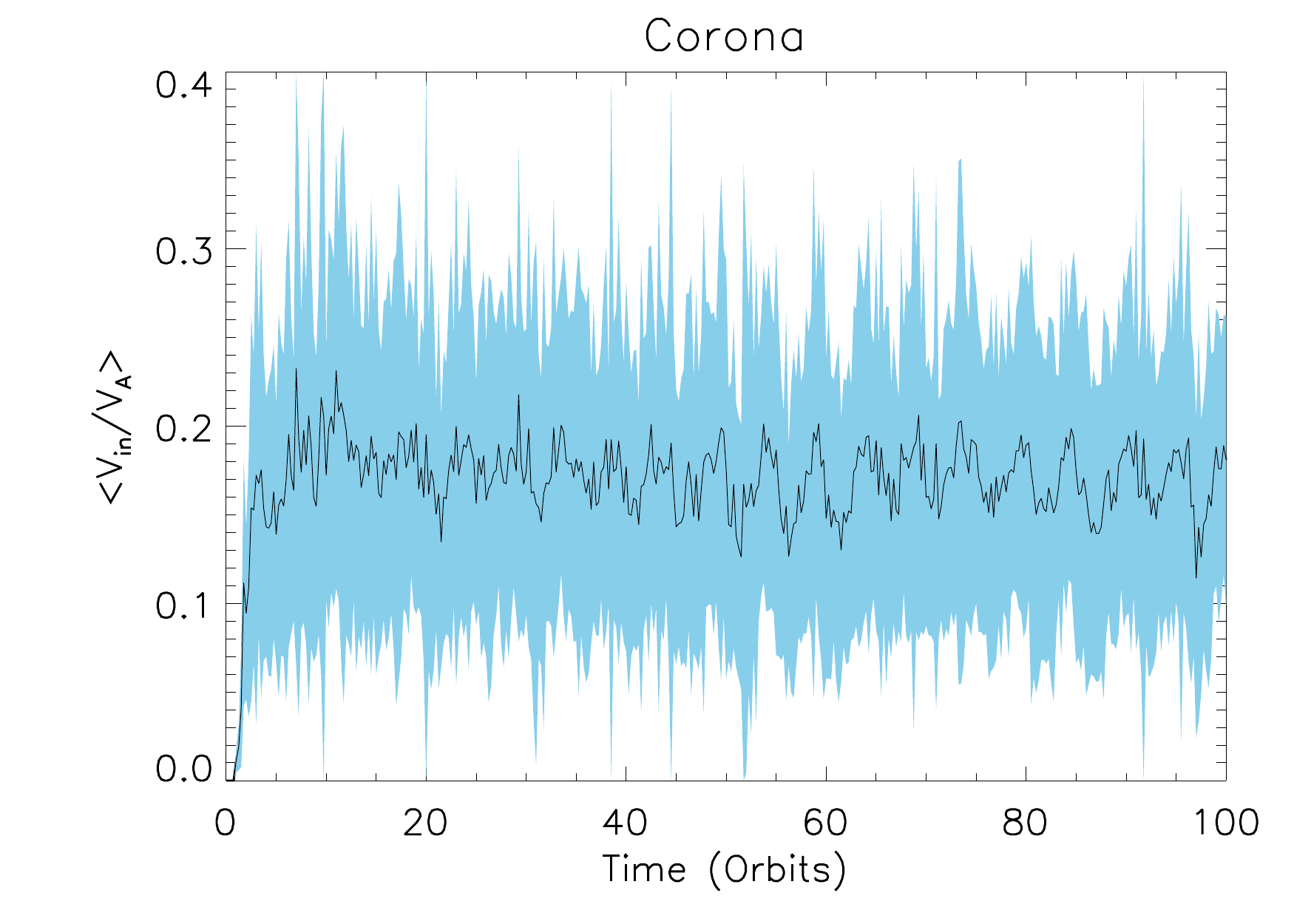}
\includegraphics[scale=0.5]{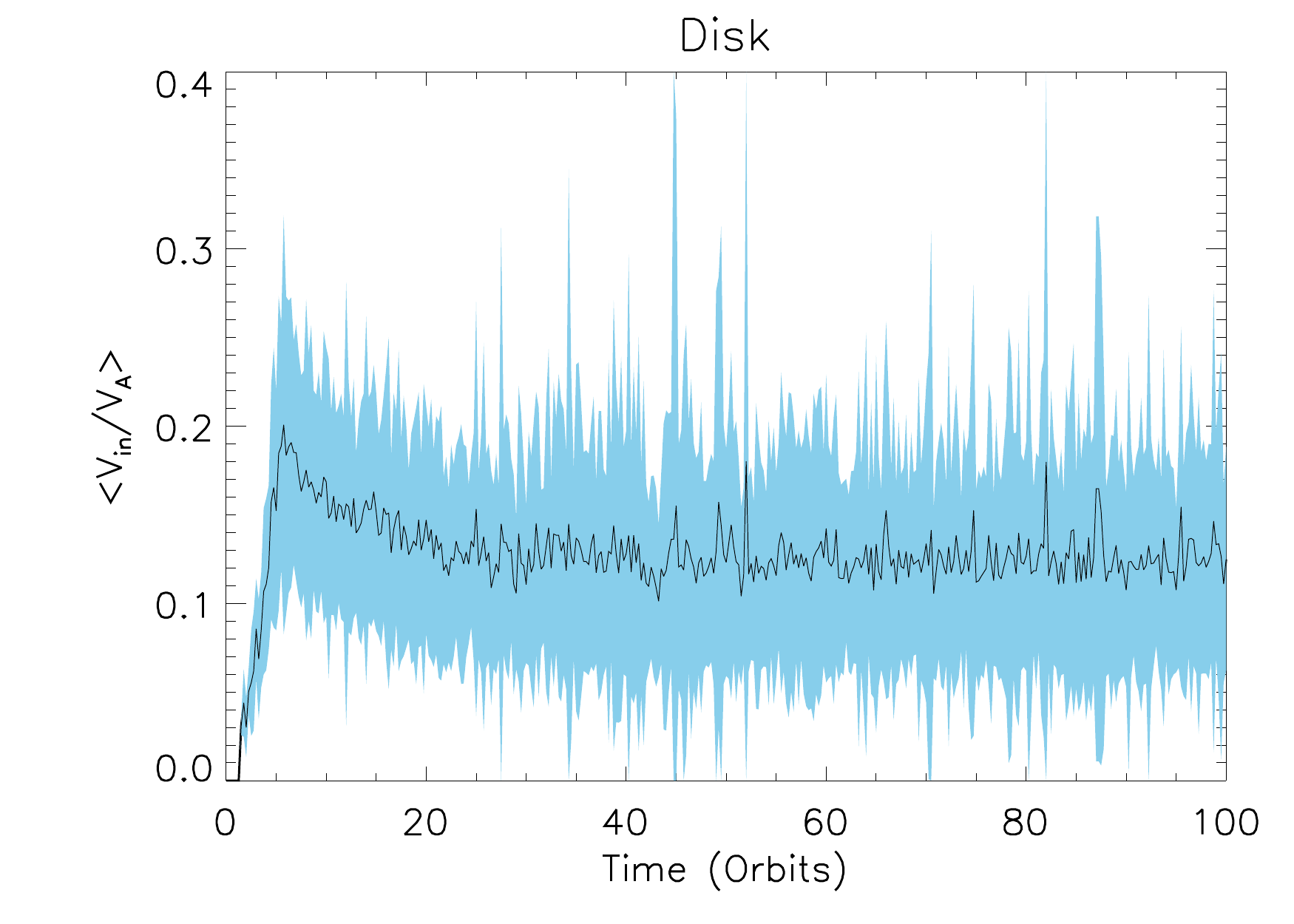}
\caption{Time evolution of the magnetic reconnection rate ($\langle V_{rec} \rangle$) of the R21b1 model evaluated in the disk and corona separately. The continuous line corresponds to the mean value obtained from the histograms for each time step and the blue shade corresponds to the standard deviation.}
\label{fig:21H_vrecevol}
\end{figure}

The distribution of $\langle V_{rec} \rangle$ was obtained for each time step of the simulations and  
Figure \ref{fig:21H_vrecevol} depict the time evolution of the average value of $V_{rec}$ for the upper and lower coronal regions together and the disk (continuous black line). These averages were calculated out of $800$ histograms computed over $100$ orbital periods. The diagrams also depict the standard deviation (blue shade) of each distribution.

In the corona, $\langle V_{rec} \rangle$  increases very fast in the early PRTI regime (at the first $3$ orbital periods)  and then saturates (after $10$ orbital periods) to a mean value varying between $\sim 0.11$ and $0.23$.
In the disk, $\langle V_{rec} \rangle$ also increases fast in the PRTI regime, achieving a peak value ($\sim 0.2$) around $\sim 6$ orbital periods, then decreases to  nearly constant value ($0.1-0.18$) after $t=20P$ when the MRI sets in, following the same trend of the time evolution of \added{the $\alpha$ parameter} and magnetic energy  (Figure \ref{fig:comp_beta}).

These values of $\langle V_{rec} \rangle$ are in agreement with the predictions of the theory of fast magnetic reconnection driven by turbulence \citep{lazarian_vishiniac_99} and the numerical studies that have tested it \citep[see, e.g.,][]{kowal_etal_09, takamoto_etal_15, singh_etal_15}.  They are also compatible with observations of fast magnetic reconnection in the solar corona associated to flares \citep[$\sim 0.001-0.1$ and up to $0.5$, see, e.g.,][]{dere_1996, aschwanden_etal_01, su_etal_2013}. 
This indicates that the algorithm used here can to be an efficient tool for the identification of magnetic reconnection sites in numerical MHD simulations of systems involving turbulence. Furthermore, it has demonstrated the ability of the PRTI and  MRI induced-turbulence to drive fast reconnection  in accretion disks and their coronae. 

\begin{figure*} 
\centering
\includegraphics[scale=0.5]{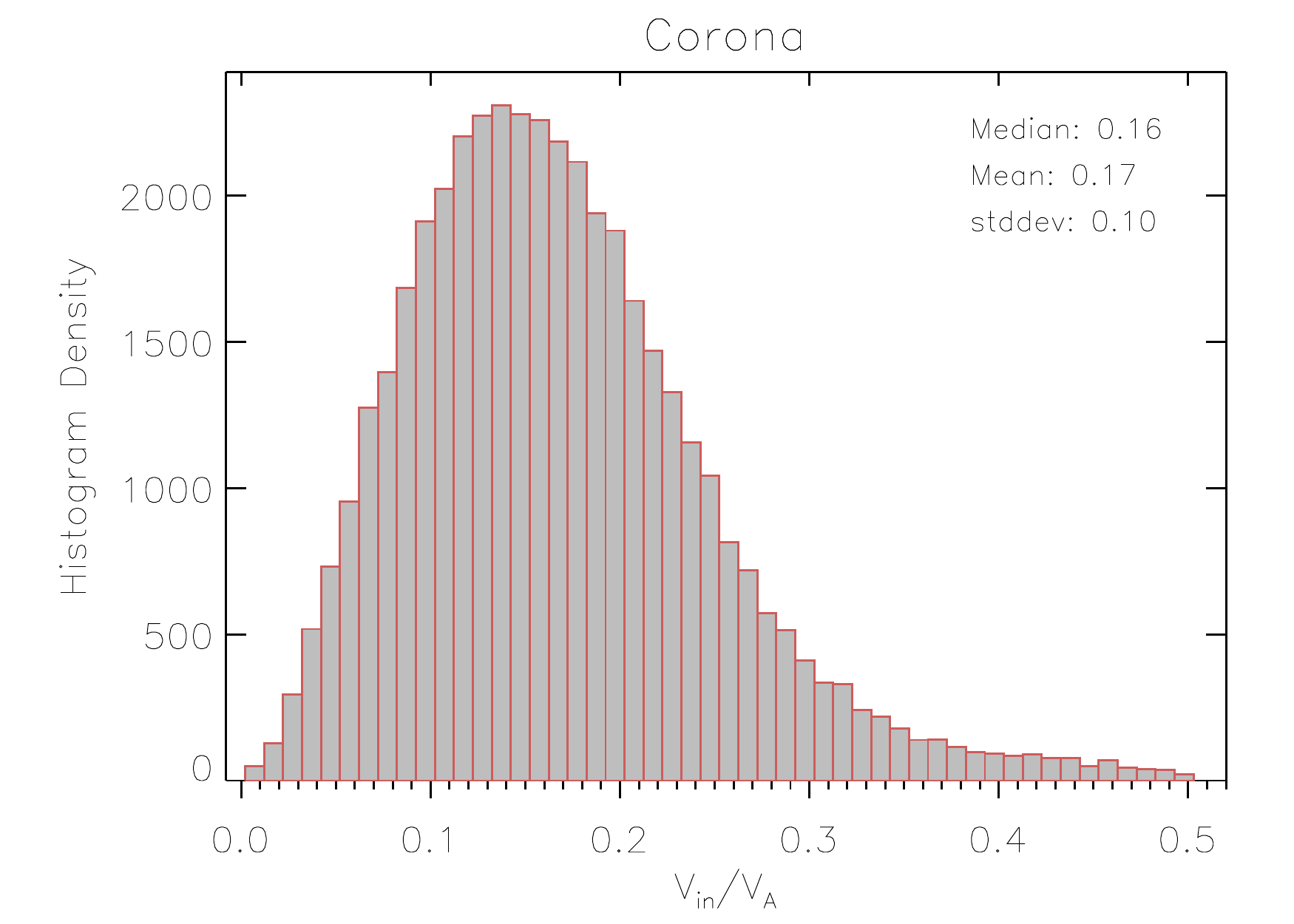}
\includegraphics[scale=0.5]{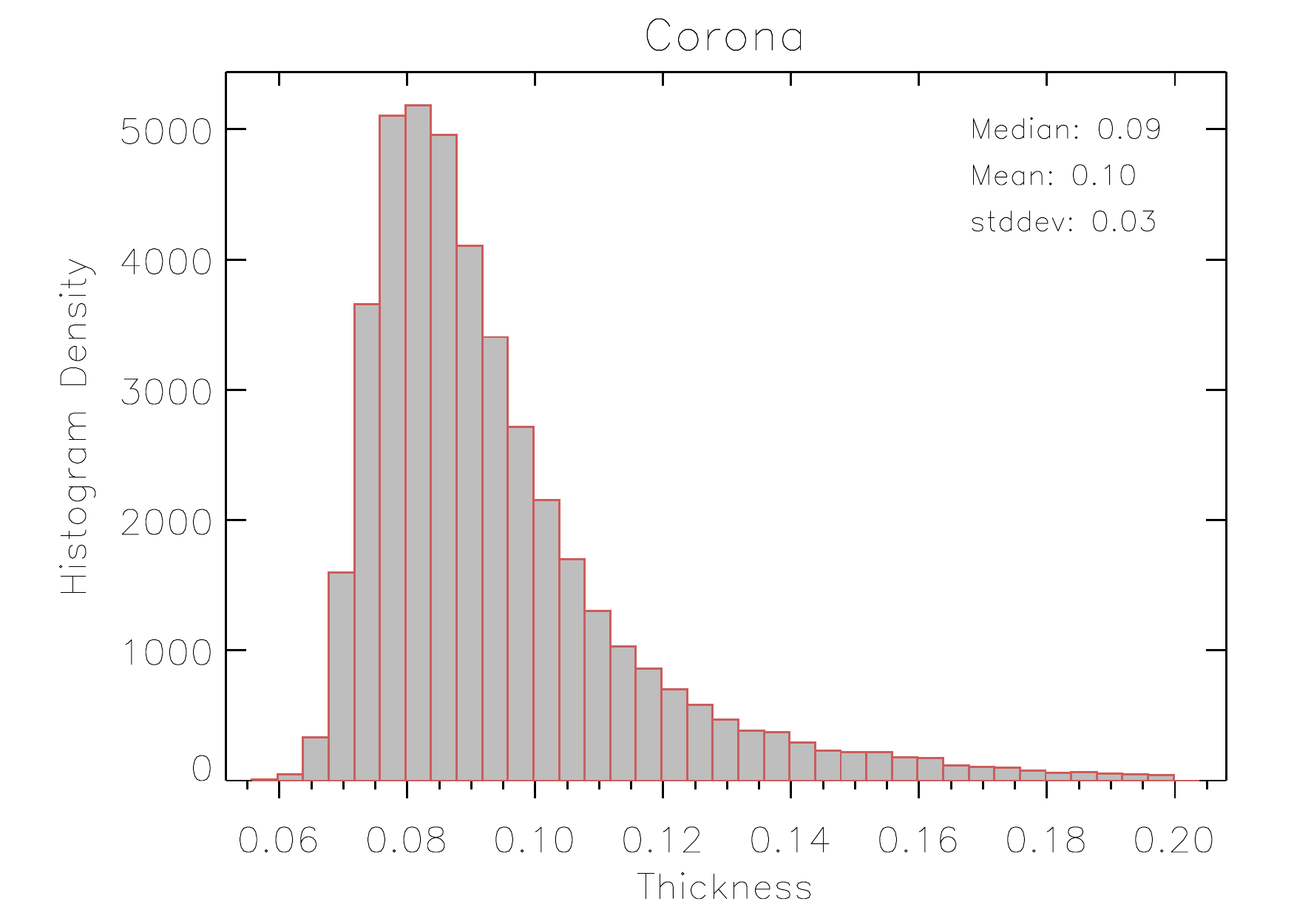}
\includegraphics[scale=0.5]{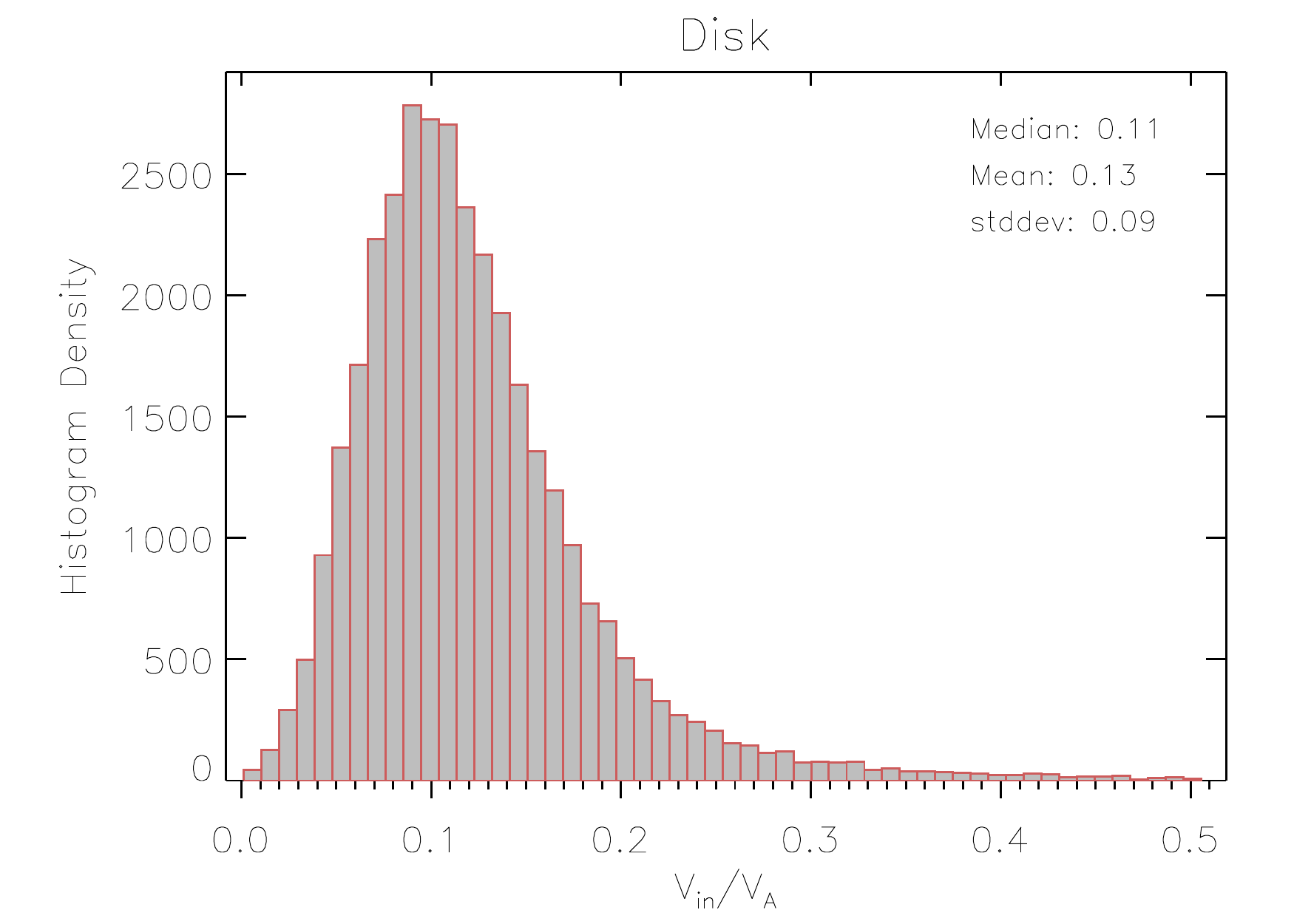}
\includegraphics[scale=0.5]{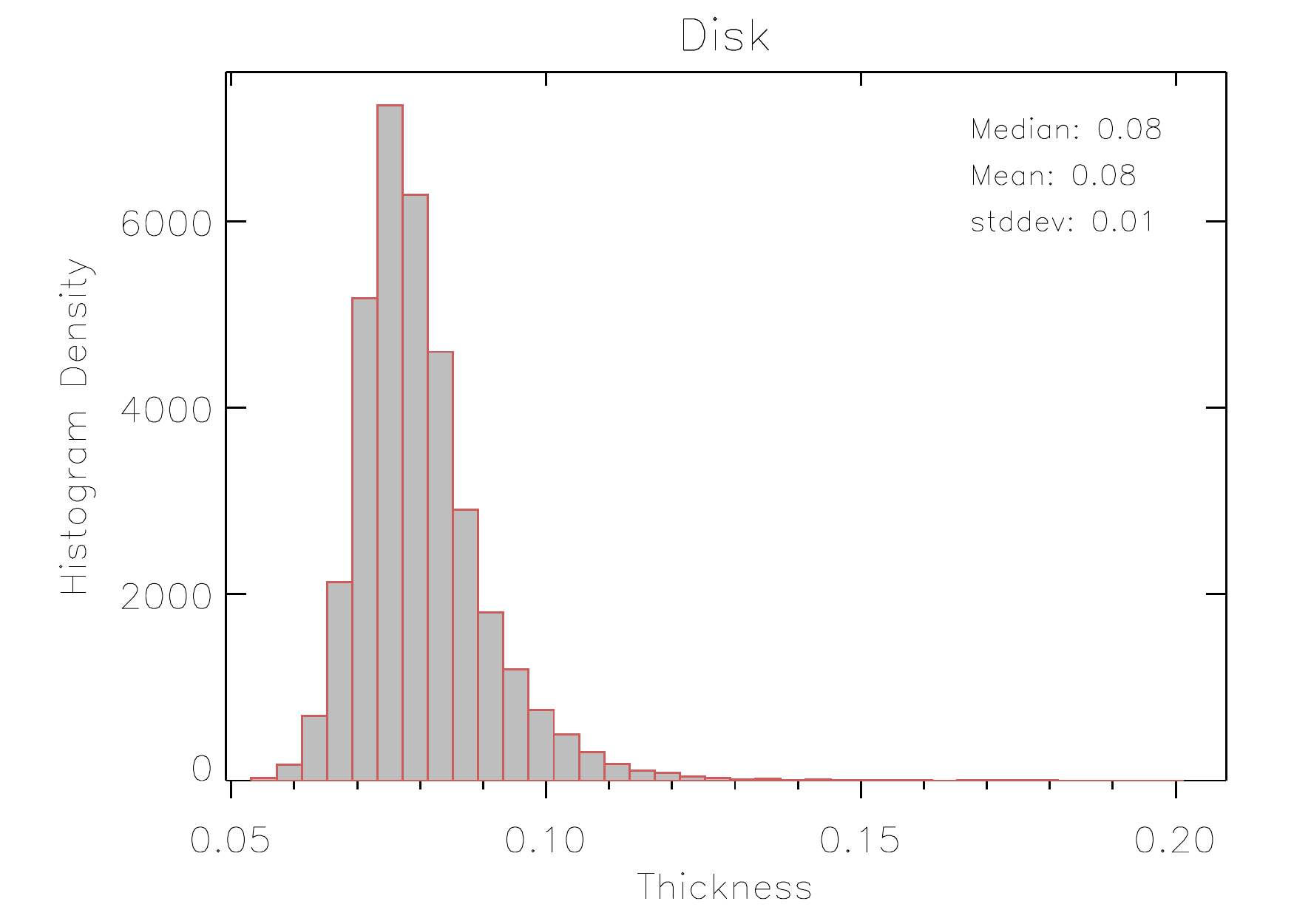}
\caption{Histograms of the magnetic reconnection rates (left diagrams) and the thickness (right diagrams) of the reconnection sites in the coronal region (upper diagrams) and disk (lower diagrams). The distributions have been obtained between $20$ and $100$ orbital periods.}
\label{fig:21H_histtot}
\end{figure*}

Figure \ref{fig:21H_histtot} shows the histograms of $V_{rec}$ (diagrams on the left side) and the measure of the thickness of the reconnection regions (diagrams on the right side) for  R21b1 model obtained between $20$ and $100$ orbital periods in the coronal region (upper diagrams) and disk (lower diagrams), where  the reconnection rate, as the other quantities, achieves a nearly steady-state. In both cases, the distributions of $V_{rec}$ and the thickness do not resemble a normal distribution, showing a long tail on the right (a skewed distribution). We should note that the algorithm has identified in the tail very few events with magnetic reconnection rates larger than $1.0$, but we have constrained the histogram to the range of  $V_{rec}$ between $0.0$ to $0.5$  which corresponds  to $99.9\%$ of the total data (similar procedure was applied for the histograms of the thickness). This may reflect the limitations of  the method to calculate the magnetic reconnection rate since the inflow velocities at the upper and lower edges in the reconnection site are not perfectly symmetric. Figure \ref{fig:21H_histtot} also depicts  for the skewed distributions both  the median (evaluated from the total data) and  the mean values with the standard deviations. For the coronal region, we have obtained an average reconnection rate of the order of $0.17 \pm 0.10 $ and a median of $\sim 0.16$, whereas in the disk we have obtained an average of \added{$0.13 \pm 0.09$} and a median of $\sim 0.11$. The averages values of the thickness show that the reconnection sites occupy $2$ or $3$ cells (for the resolution of $21H^{-1}$), therefore numerical effects could affect the evaluation of $V_{rec}$. We have also performed tests with larger reconnection sites, but the averages of $V_{rec}$ did not change significantly.

\added{As mentioned in section \ref{sec:met_vrec}, we have constrained our sample  in the analysis above considering  only symmetric profiles of the velocity and magnetic fields around the reconnection sites. Here, we repeated the same analysis using the whole sample (imposing no restriction) and another one including both non-symmetric and symmetric profiles. In both cases, the mean values and standard deviations are higher than the ones shown previously (in Figure \ref{fig:21H_vrecevol}), as we expected since $V_{rec}$ has been evaluated through equation (\ref{eq:vrec}), where we averaged the ratio between the inflow velocity and Alfv\'{e}n speed at the top and the bottom of the reconnection site. When considering the whole sample, we have obtained $V_{rec}^{corona} = 0.24 \pm 0.24$ and $V_{rec}^{disk} = 0.20 \pm 0.25$, while  for the sample including both symmetric and nonsymmetric profiles we obtained $V_{rec}^{corona} = 0.17 \pm 0.13$ and $V_{rec}^{disk} = 0.14 \pm 0.15$. With regard to  the thickness of the reconnection regions in these cases, the average values have not changed with respect to  the previous results (Figure \ref{fig:21H_histtot}). This is not a surprise since the criteria to evaluate the thickness are independent of the symmetry of the velocity and magnetic profiles.}

\subsection{Correlation with turbulence} 

In order to verify more quantitatively the correlation between the magnetic reconnection rate and the turbulence, we have performed a Fourier analysis and evaluated the two-dimensional power spectrum of the velocity field\footnote{The power spectrum was obtained from the velocity field without the advection term ``$-{3\over 2}\Omega_{0} x~\widehat{\boldsymbol{y}}$'', computed by the fargo scheme (see section \ref{sec:num_met}).} ($P(k,z)=|\boldsymbol{\widetilde{u}}(k_x, k_y,z)|^2$) for the $k_{x}$ and $k_{y}$ wavenumbers in different heights (``$z$'' direction) inside the computational domain. The power spectrum has been taken from the average in \added{annular areas between $k-dk$ and $k$ (where $k=\sqrt(k_x^2+k_y^2)$)} in the $k_x-k_y$ Fourier space, as described in section \ref{sec:Diagnostics} (see eq.\ref{eq:power_spectrum}). We have also averaged in time with bins of $1$ orbital period to reduce the fluctuations of the spectra. \added{Such decomposition has been chosen in order to verify the turbulence level in the disk and coronal regions separately, since the diagrams of $V_{rec}$ (see Figure \ref{fig:21H_vrecevol}) have shown significant differences between these two regions after $t=20P$.}

\begin{figure*}
\includegraphics[scale=0.55]{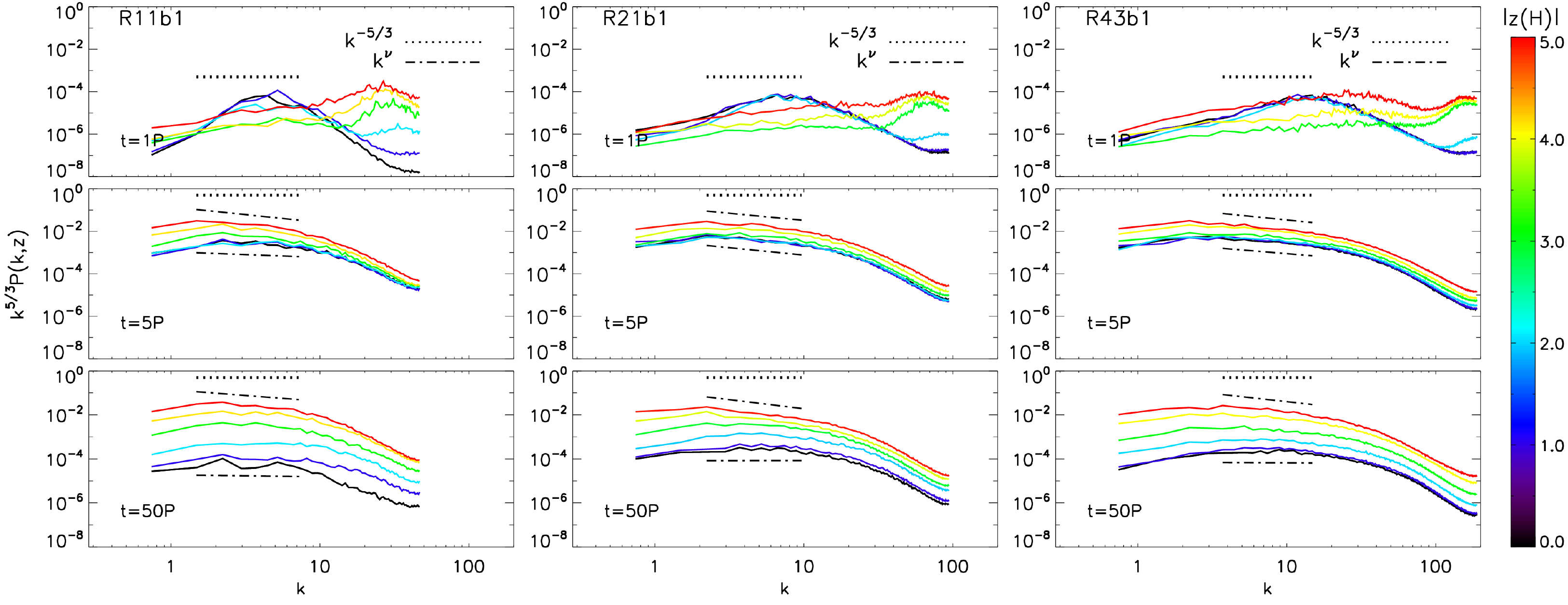}
\caption{Compensated power spectrum of the velocity field ``$k^{5/3} P(k,z)$'' taken at $t=1P$, $5P$ and $50P$ (from the top to the bottom) \added{for the models R11b1 (left row), R21b1 (central row),  and R43b1 (right)}. The colors indicate the height where the power spectrum have been evaluated. The dashed line corresponds to the Kolmogorov spectrum $k^{-5/3}$ \added{and the dot-dashed line corresponds to the spectrum $k^{\nu}$ fitted for each curve}.}
\label{fig:21H_fft}
\end{figure*}

Figure \ref{fig:21H_fft} shows the power spectrum \added{for the low, intermediate and high-resolution simulations (R11b1, R21b1, and R43b1 models, respectively)} compensated by a factor of $k^{5/3}$ in three different times, at $t=1P$, $5P$ and $50P$, where the colors of each line correspond to the height ``$z$''. As we expected, the turbulence is not fully developed in the early stages of the simulations. On the other hand, after $5$ orbital periods the turbulent power spectrum is much larger and shows the typical cascading to small scales with \added{a slope ($k^{\nu}$) between $-1.9 < \nu < -2.4$ in the coronal region and disk (for all the resolutions). At $t=50P$, the slope in the middle plane of the disk ($z=0$) is $\nu \sim -1.7$ (i.e., Kolmogorov like) and increases as a function of the altitude, reaching a value of $\nu \sim -2.5$ at the coronal region which is closer to an $-8/3$ power law, typical of a 2D turbulent spectrum distribution (eq.\ref{eq:power_spectrum}), suggesting a change of regimes as we go from the disk to the corona, i.e., from large to small values of $\beta$. We should remember that even at these evolved stages, the corona keeps traces of large scale coherent magnetic loops embedded in the turbulent flow, developed during the PRTI regime. No significant differences are seen when considering different resolutions.\footnote{The behavior of the power spectrum for the MRI (and PRTI)-driven turbulence is far from being understood yet. For instance, recently \cite{walker_etal_16}, considering shearing-box simulations, have decomposed the energy power spectrum into a parallel (large-scale shear-aligned) and a perpendicular (small-scale fluctuation) component to the mean magnetic field. With this procedure, they obtained power law spectra close to $k^{-2}$ and $k^{-3/2}$ for the parallel and perpendicular components, respectively. A similar decomposition is out of the scope of the present work, but their results for the small scale, turbulent fluctuations are  compatible with our results.}}

At $t=1P$, for the R21b1 model (second diagram of Figure \ref{fig:21H_fft}),   the power is still very small and the magnetic reconnection rate has values below $0.1$ (see Figure \ref{fig:21H_vrecevol}).
Around $t=5P$, during the PRTI, the magnetic reconnection rate achieves its largest values between $0.1$ in the disk and $0.2$ in the coronal region. The power in the middle of the disk is smaller than in the coronal region and is consistent with the behavior of $\langle V_{rec} \rangle$ in such regions (where the average value in the coronal region is higher than in the disk after $20$ orbital periods). Similar behavior is found at $t=50P$ indicating a correlation between the turbulence and $V_{rec}$.

\added{Figure \ref{fig:21H_fft} also shows that the low-resolution model R11b1, despite the different behaviour in Figure  \ref{fig:comp_11H21H43H} that indicates a decrease of the volume-averaged magnetic energy density with time,  has a similar  velocity field  power spectrum to the higher resolution models R21b1 and R43b1. This similarity between the models of different resolution will be further stressed in the next subsection.}

\subsection{Comparison between models}
\label{sec:comp_vrec}

\begin{figure*}
\centering
\includegraphics[scale=0.5]{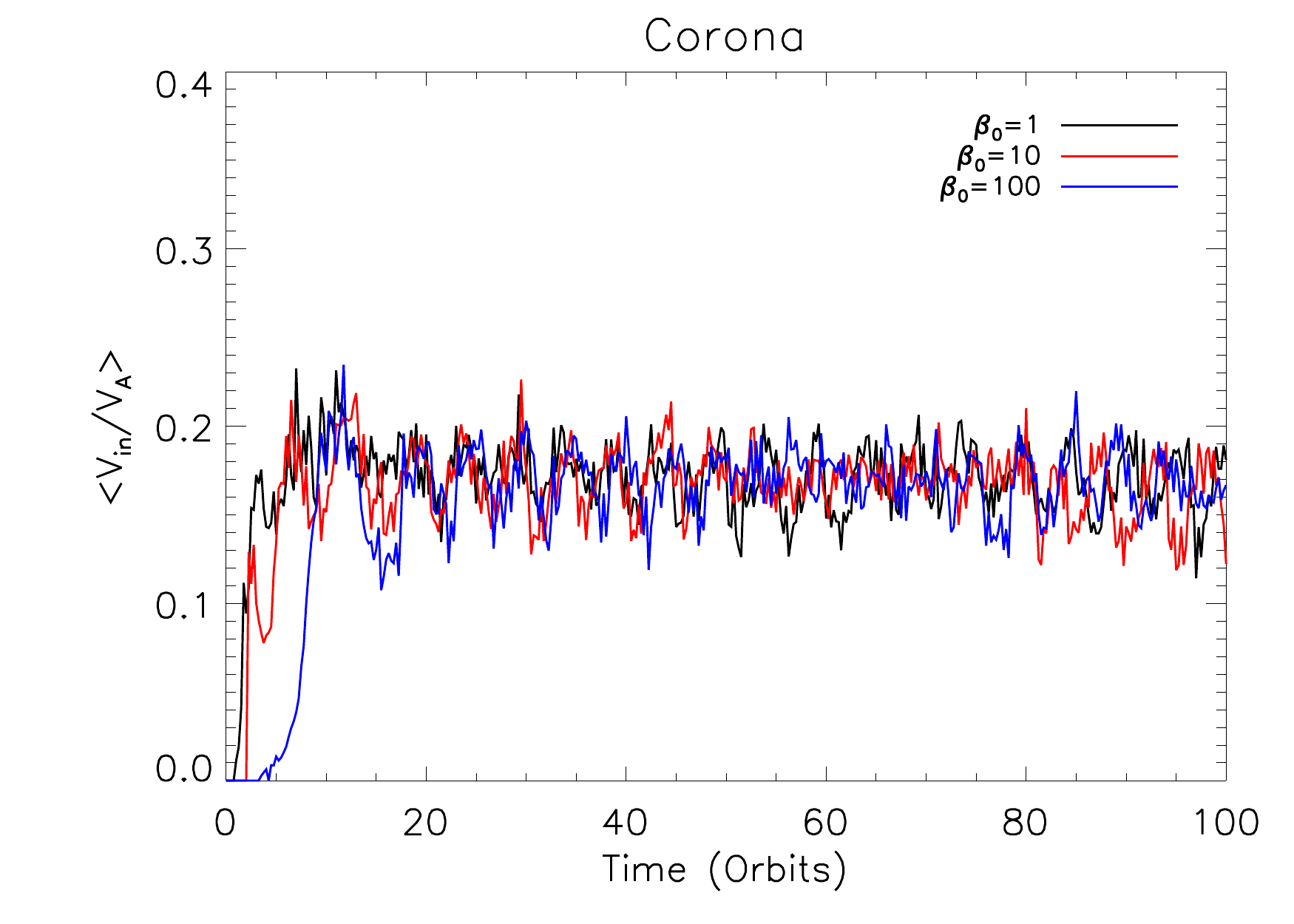}
\includegraphics[scale=0.5]{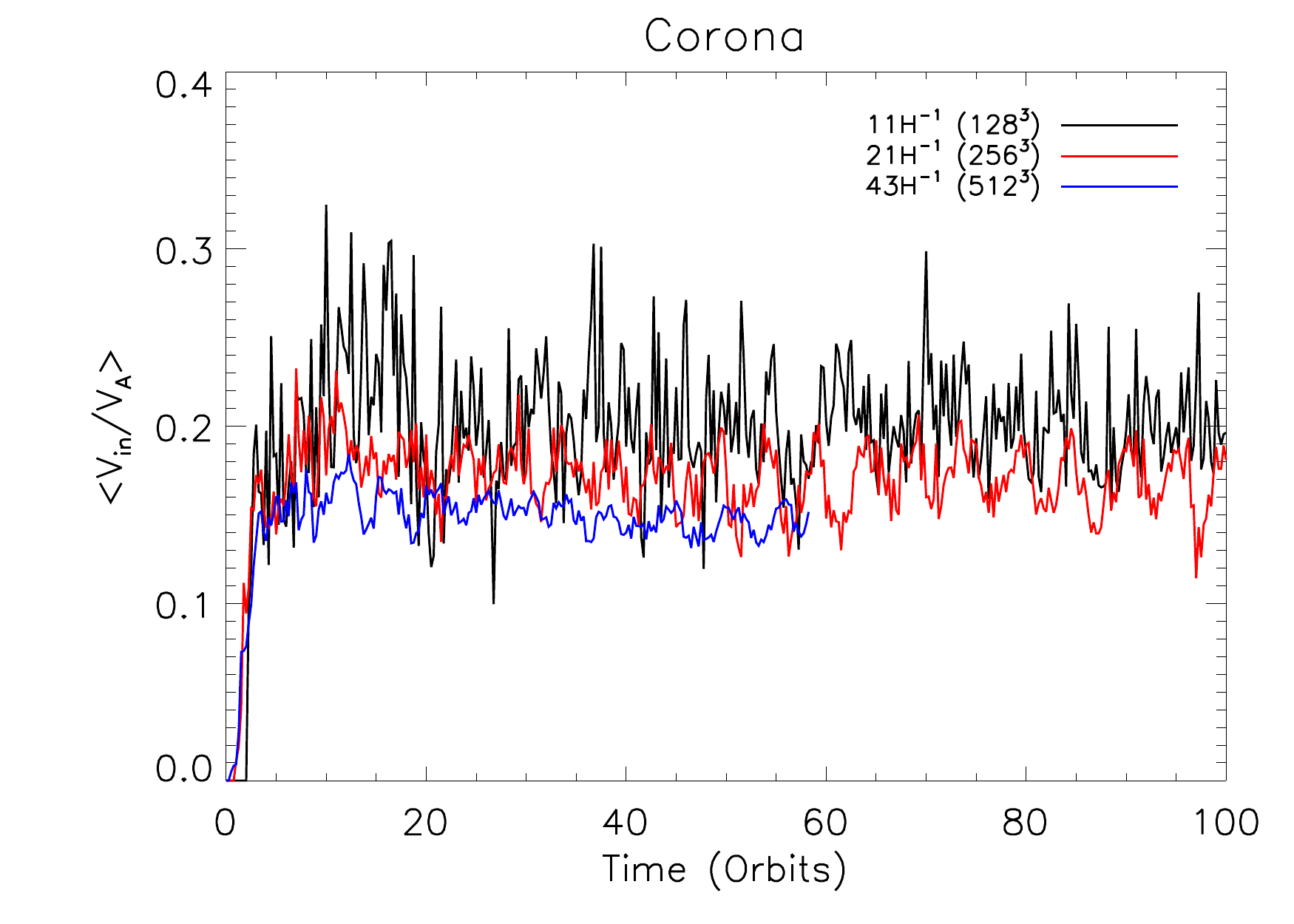}
\includegraphics[scale=0.5]{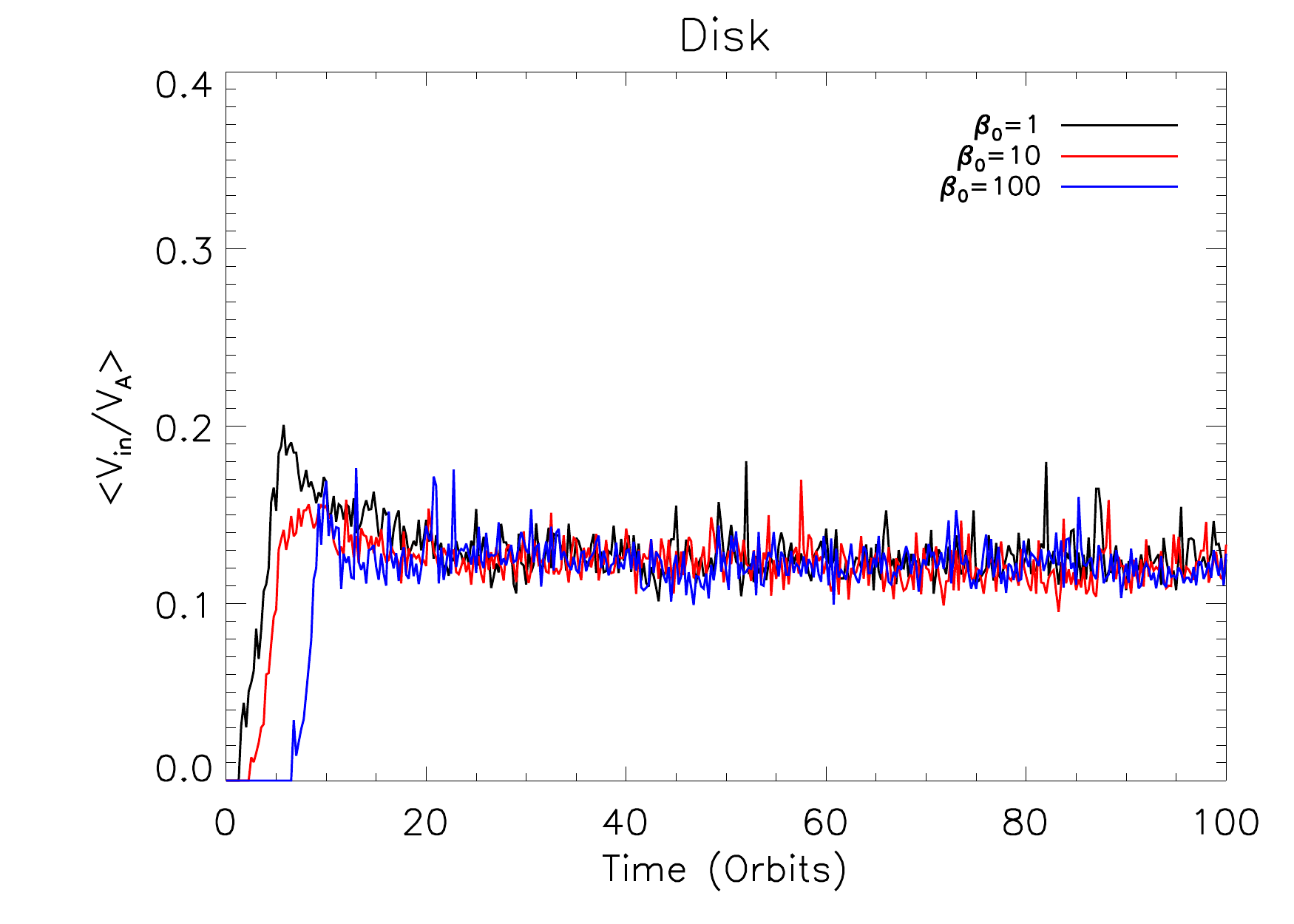}
\includegraphics[scale=0.5]{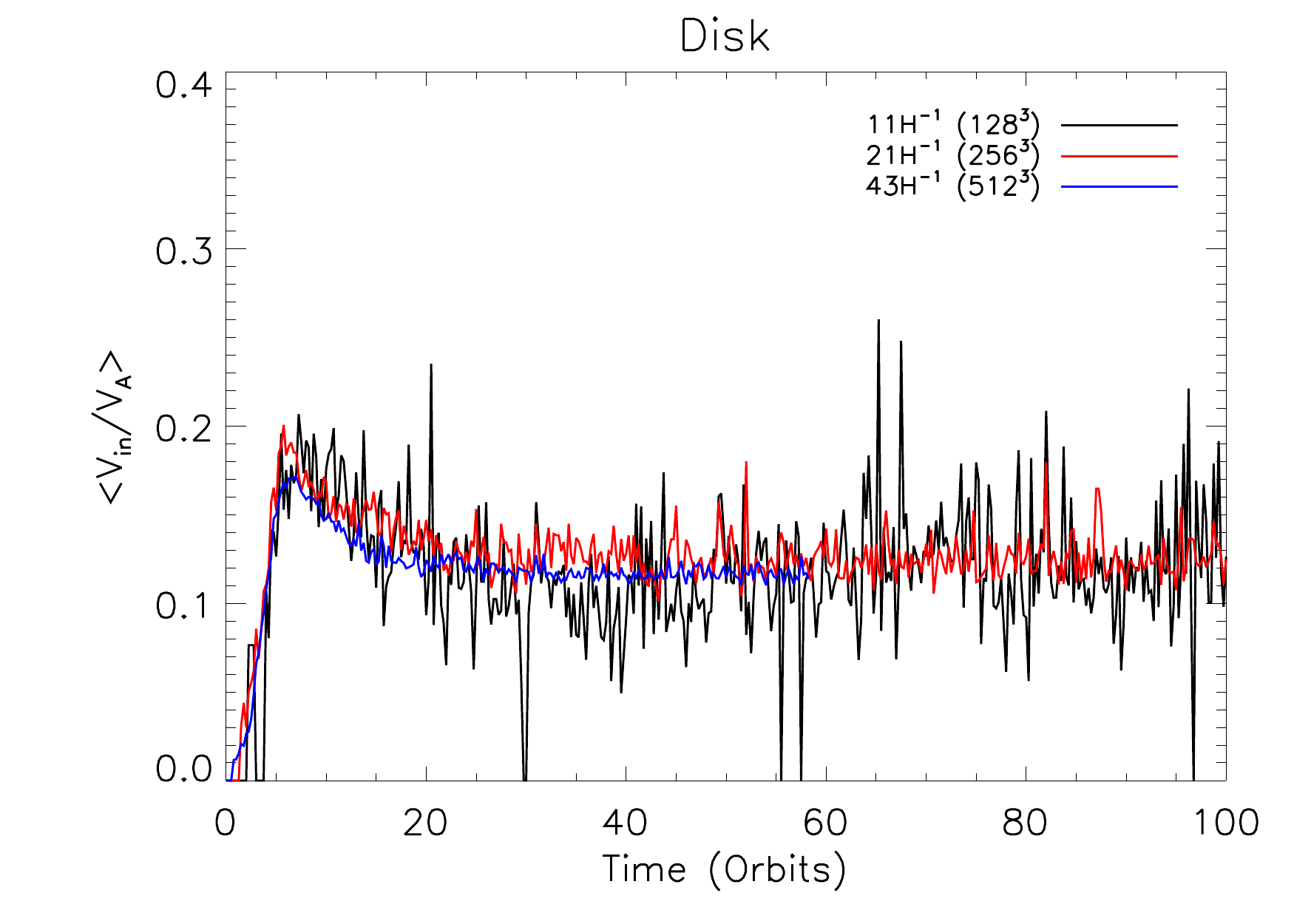}
\caption{The left diagrams show the time evolution of the magnetic reconnection rate $\langle V_{rec} \rangle$ for different initial values of $\beta_{0}$ (for the $21H^{-1}$ resolution). The black line corresponds to the reference model with $\beta_{0}=1$ (R21b1), whereas the red and blue lines correspond to the models with $\beta_{0}=10$ (R21b10) and $100$ (R21b100), respectively. The right diagrams show the time evolution of $\langle V_{rec} \rangle$ for different resolutions. In this case, the black line corresponds to the low resolution model ($11H^{-1}$, R11b1), whereas the red and blue lines correspond to the intermediate ($21H^{-1}$, R21b1) and high ($43H^{-1}$, R43b1) resolution simulations, respectively. We have evaluated $\langle V_{rec} \rangle$ in the coronal region (upper diagram) and in the disk (lower diagram).}
\label{fig:comp_vrecbeta}
\end{figure*}

Figure \ref{fig:comp_vrecbeta} compares the time evolution of $\langle V_{rec} \rangle$ for simulations with different values of $\beta_{0}$ (models R21b1, R21b10 and R21b100, left diagrams) and resolutions (models R11b1, R21b1 and R43b1, right diagrams). Despite the high standard deviations (see, e.g., Figure \ref{fig:21H_vrecevol}), the left diagrams show that the behavior of $\langle V_{rec} \rangle$ is similar to the one of the reference model R21b1, and consistent with the time evolution of these systems as discussed in Section \ref{sec:comp_beta}. In the first orbital periods (before $20P$), in the PRTI regime, the models with different $\beta_{0}$ show an increase of the reconnection rate, but with a significant delay for those with higher $\beta_{0}$. Above $20$ orbital periods, regardless of the initial $\beta_{0}$, $\langle V_{rec} \rangle$ converges to the average values discussed previously (between $0.11$ and $0.23$). At this stage, the volume averages of the magnetic energy density \added{and $\alpha$} also converge to a steady-state (as seen in Figure \ref{fig:comp_beta}). 
The comparison of $\langle V_{rec} \rangle$ with different resolutions reveals a good convergence mainly in the disk region (bottom diagram in the right side of Figure \ref{fig:comp_vrecbeta}), whereas in the coronal region (top diagram in the right side) the mean values of $V_{rec}$ for the high-resolution models are slightly smaller than for the low-resolution.

Table \ref{tab:vrec} summarizes the time average values of the magnetic reconnection rate $\langle V_{rec} \rangle_{t}$ \added{, the thickness $\langle \delta \rangle_{t}$, and the number $\langle N \rangle_{t}$ of reconnection sites identified by the algorithm} obtained between $20$ and $100$ orbital periods, excepted for the high-resolution simulation (model R43b1) whose time average was obtained between $20$ and $58$ orbital periods. Considering the standard deviations, all the models show $\langle V_{rec} \rangle_{t}$ values which are compatible, both for different values of $\beta$ and resolutions. 

\added{The averaged thickness $\langle \delta \rangle_{t}$ for the high-resolution model is significantly smaller than that of the intermediate and low-resolution models, as we expect, since the structures of the magnetic reconnection sites are better resolved for the R43b1 model. Besides, $\langle V_{rec} \rangle_{t}$ is not strongly affected,  neither by the thickness nor by the numerical resistivity\footnote{In this work, as described at the beginning of this section, we have not applied an explicit resistivity in the simulations, thus the magnetic reconnection process occurs in presence of numerical resistivity. In this case, the role of the resistivity can be evaluated by comparing the models with the three different resolutions employed here ($11H^{-1}$, $21H^{-1}$, and $43H^{-1}$).} for different resolutions. This is not a surprise since, according to the turbulence-induced fast reconnection theory \citep[see][]{lazarian_vishiniac_99, eyink_etal_11, eyink_etal_13}, the presence of turbulence speeds up the reconnection independently of the Ohmic resistivity (here mimicked by the numerical resistivity, as stressed before; see also eq.\ref{eq:vrec_turb}).}  

\added{Finally, as the resolution increases, the number of identified reconnection sites increases as a consequence of the better-resolved magnetic structures, which improve the statistical analysis of $V_{rec}$. For this reason, the right diagrams of Figure \ref{fig:comp_vrecbeta} show that the amplitude variability of $\langle V_{rec} \rangle$ decreases for the models with higher resolution. Above $20$ orbital periods, the model R43b1 shows $\langle V_{rec} \rangle$ values varying between $\sim 0.13$ and $0.17$ in the coronal region, and between $\sim 0.11$ and $0.13$ in the disk. On the other hand, the model R11b1 shows $\langle V_{rec} \rangle$ values between $\sim 0.10$ and $0.30$ in the coronal region, and between $\sim 0.0$ and $0.26$ in the disk. Despite the increase in the number of identified reconnection sites, the distributions in time seen in Figure \ref{fig:21H_histdb} do not change significantly for different resolutions.}

\begin{table*}
\centering
\caption{Time average of $V_{rec}$, the thickness ($\delta$), and the number of identified reconnection sites ($N$).}
\begin{tabular}{lccccccc}
\hline \hline \\[-2ex]
\multicolumn{1}{l}{Simulation} &
\multicolumn{2}{c}{$\langle V_{rec} \rangle_{t} \pm \sigma$} &
\multicolumn{2}{c}{$\langle \delta \rangle_{t} \pm \sigma$} &
\multicolumn{2}{c}{$\langle N \rangle_{t} \pm \sigma$} &
\multicolumn{1}{c}{$\Delta T$}

\\

\multicolumn{1}{l}{Name} &
\multicolumn{1}{c}{Corona} &
\multicolumn{1}{c}{Disk} &
\multicolumn{1}{c}{Corona} &
\multicolumn{1}{c}{Disk} &
\multicolumn{1}{c}{Corona} &
\multicolumn{1}{c}{Disk} &
\multicolumn{1}{c}{Range}
\\[0.5ex] \hline
\\[-1.8ex]
R11b1   & $0.20 \pm 0.10$ & $0.12 \pm 0.07$ & $0.17 \pm 0.04$ & $0.15 \pm 0.02$ & $25 \pm 9$ & $10 \pm 5$ & 20P-100P \\
R21b1   & $0.17 \pm 0.10$ & $0.13 \pm 0.09$ & $0.10 \pm 0.03$ & $0.08 \pm 0.01$ & $134 \pm 34$ & $107 \pm 16$ & 20P-100P \\
R21b1\footnote{Sample with the whole data (without restrictions, see \ref{sec:vrec}) .}& $0.24 \pm 0.24$ & $0.20 \pm 0.25$ & $0.10 \pm 0.03$ & $0.08 \pm 0.01$ & $559 \pm 74$ & $402 \pm 42$ & 20P-100P \\
R21b1\footnote{Sample with symmetric and nonsymmetric profiles (see \ref{sec:vrec}).}& $0.17 \pm 0.13$ & $0.14 \pm 0.15$ & $0.10 \pm 0.03$ & $0.08 \pm 0.01$ & $216 \pm 44$ & $163 \pm 21$ & 20P-100P \\
R21b10  & $0.17 \pm 0.10$ & $0.12 \pm 0.09$ & $0.09 \pm 0.03$ & $0.08 \pm 0.01$ & $137 \pm 34$ & $106 \pm 15$ & 20P-100P \\
R21b100 & $0.17 \pm 0.10$ & $0.12 \pm 0.08$ & $0.10 \pm 0.03$ & $0.08 \pm 0.01$ & $135 \pm 33$ & $105 \pm 15$ & 20P-100P \\
R43b1   & $0.15 \pm 0.09$ & $0.12 \pm 0.08$ & $0.05 \pm 0.01$ & $0.04 \pm 0.01$ & $726 \pm 103$ & $675 \pm 50$ & 20P-58P 
\\[2ex]
\hline
\label{tab:vrec}
\end{tabular}
\end{table*}

\section{Discussion and Conclusions}
\label{sec:discussion_conclusion}

In this work, we have performed 3D-MHD shearing-box numerical simulations \citep[see][]{hawley_etal_95} of accretion disks in order to capture with a resolution as high as possible, the long term dynamical evolution \added{of the system,} where Parker-Rayleigh-Taylor and magnetorotational instabilities (PRTI and MRI, respectively) develop, and follow the formation of the disk corona and turbulence. \added{Our main goal here was to understand the development of fast magnetic reconnection in accretion disk/corona systems induced by turbulence. As stressed in Section \ref{sec:intro}, magnetic reconnection events have an important role on heating  and acceleration of  particles in the plasma.}

The present study is applicable to accretion phenomena in general, but may be particularly relevant for accretion disks around stellar mass and supermassive black holes which are believed to sustain strong magnetic fields, at least during certain accretion regimes, which could be produced by dynamo processes and/or by the transport of magnetic fields from a companion star or the surrounding medium \citep[see][]{dgdp_lazarian_05, dgdp_etal_10a, kadowaki_etal_15, singh_etal_15}.

In our simulations, in order to allow for the growth of the PRTI and MRI, we have considered accretion disks with an initial strong azimuthal magnetic field, having  $\beta_{0} =$ $1$, $10$ and $100$.

\added{In the following we summarize our main results and compare them with previous works when applicable:} 

\begin{itemize}

\item
\added{As expected,} the PRTI, which dominates the early evolution of the system due to the small values of $\beta$, leads to the formation of poloidal magnetic fields and loops which are transported from the midplane of the disk to the higher \added{altitudes}, allowing for the formation of a magnetized corona with complex structure and $\beta$ values around unit. The increase of $\beta$ in the disk, on the other hand, allows for the development of the MRI (which becomes dominant after $\sim 20$ orbital periods in all models), and both instabilities drive turbulence and a dynamo action in the system, \added{in agreement with previous works \citep[see, e.g.,][]{johnson_etal_2008,salvesen_etal_16b,salvesen_etal_16a}.}

\item
We have employed outflow boundaries conditions in the vertical direction which are more suitable to reproduce real accretion disk coronae. These boundaries allow for the transport of magnetic field to outside of the computational domain that causes the decay of the total magnetic field with time until it achieves a nearly steady-state regime, this due to the generation of new field flux by dynamo process mainly during the MRI regime. \added{We should notice that earlier works have explored the role of different boundaries in the vertical direction \citep[see, e.g.,][]{salvesen_etal_16a} and, though out of the scope of this work, we  have also performed simulations with periodic boundary conditions  obtaining  similar results to those of these authors.} 

\item
 Our systems end up with a gas pressure-dominated disk (with $\beta>>1$) and a magnetically-dominated corona (whith $\beta \simeq 1$)  which is consistent with the results found in \cite{miller_stone_2000, salvesen_etal_16a}. This behavior is not in agreement with the results of \cite{johansen_levin_08}, who obtain a highly  magnetized disk. This is probably due to their adopted vertical boundary conditions which prevent the escape of the azimuthal magnetic flux from the domain. In a more recent work, \cite{salvesen_etal_16b, salvesen_etal_16a}, considering similar boundary conditions (with initial zero net vertical magnetic flux) as in here, have also found that strong azimuthal fields cannot be maintained for long periods \textit{within} the disk due to the magnetic buoyancy effects, unless the system has sufficient initial net vertical flux.

\item
We have tested the numerical convergence of our results with different resolutions ($11H^{-1}$, $21H^{-1}$ and $43H^{-1}$) and found a good agreement between the intermediate (our reference model, R21b1) and high (R43b1) resolution models which reach the steady-state regime at the same time when the MRI becomes self-sustained inside the system. The low-resolution model (R11b1), on the other hand, shows a continuous slow temporal decrease of the magnetic field after $20$ orbital periods without achieving a steady-state indicating that this model did not achieve appropriate resolution. 

\item
\added{Though not in the main scope of this work, it is important to highlight the role of the PRTI as a transient phase of the accretion disk system and a natural way to increase the value of the $\alpha$-parameter in our simulations.} The values found for \added{this parameter} in the steady-state regime ($\langle\overline{\alpha}\rangle_{mag} \sim 0.02$) are smaller than those expected from observations \added{\citep[$\alpha \sim 0.1-1$, see, e.g., ][]{king_etal_07, zhu_etal_07}}. On the other hand, earlier numerical studies of homogeneous and stratified systems that have imposed initial zero net vertical fields to trigger the MRI \citep[see][]{hawley_etal_95, stone_etal_96, fromang_stone_09, davis_etal_10},  rather than an initial azimuthal magnetic field as in here, \added{obtained compatible values of $\alpha$}. 
We obtained a large \added{$\langle\overline{\alpha}\rangle_{mag} \sim 0.3$} only during the PRTI phase, in the first $10$ orbital periods. \cite{bai_stone_13} have demonstrated that a net vertical magnetic field applied to the simulations (generating stronger large-scale fields) can increase the values of $\alpha$ and explain the observational estimates discussed above, although they do not explain how these fields could arise naturally. In contrast, in our simulations \added{and previous works \citep[see][]{johnson_etal_2008,salvesen_etal_16b, salvesen_etal_16a}}, the presence of the PRTI induces both the formation of poloidal fields and the increase of \added{$\alpha$ parameter} to the observed values. We can speculate that this short period in which the PRTI grows and large-scale poloidal fields develop could be related with flare events in accretions disk systems \citep[as argued, e.g., in ][]{dgdp_lazarian_05, dgdp_etal_10a, kadowaki_etal_15, singh_etal_15}.

\item
Finally \added{and most important result of this work,} the arising of magnetic loops due to the PRTI followed by the development of turbulence due both to the PRTI and MRI produce current density peaks in the coronal region and disk, indicating the presence of magnetic reconnection.  
To track this process, we employed a modified version of the algorithm developed by \cite{zhdankin_etal_13} to identify current sheets (with strong current density) produced by the encounter of magnetic field lines of opposite polarity in the turbulent regions of the computational domain. From this analysis, we evaluated the magnetic reconnection rates employing  the method adopted by \cite{kowal_etal_09}. Despite the high standard deviations derived from the method, we have found peak values for the reconnection rate ($V_{rec}=V_{inflow}/V_{A}$) of the order of $0.2$, and average values of the order of \added{$0.13 \pm 0.09$} in the accretion disk and $0.17 \pm 0.10 $ in the coronal region 
(for our reference model, R21b1), 
indicating the presence of fast magnetic reconnection events, as predicted by the theory of turbulence-induced fast reconnection of \cite{lazarian_vishiniac_99}. Regarding the histograms of $V_{rec}$, they do not resemble a normal distribution as they exhibit a tail at higher velocities. This is probably due to the limitations of the method employed which evaluates the reconnection rate only at two points at the edges of the magnetic reconnection site along the axis of the fastest decaying of the current density (obtained from the Hessian matrix). In future work, we intend to apply different methods to compare with the current results. \cite{kowal_etal_09}, for instance, besides evaluating the magnetic reconnection rate with the same method used here, tried also another one considering the time derivative of the magnetic flux\footnote{\cite{zhdankin_etal_13} have also identified regions considering changes in the magnetic flux function, but from saddle points in a slice of the computational domain.}. However,  this is hard to apply in our simulations since the magnetic reconnection sites can move in space or change the direction with time \citep[other methods to  be considered include, e.g.,][]{greco_etal_2008, servidio_etal_2011}. 

\end{itemize}

Despite the limitations of the method, \added{as stressed in Section \ref{sec:intro},} the observations of flares in the solar corona indicate magnetic reconnection rates in a range between $0.001-0.5$ \cite[see, e.g.,][]{dere_1996, aschwanden_etal_01, su_etal_2013} and strengthen our results, \added{since the fast reconnection mechanism should be similar in most turbulent astrophysical environments, specially in coronal plasmas}. Numerical simulations of turbulent environments point to similar reconnection rates \citep[see, e.g.,][]{kowal_etal_09, takamoto_etal_15, singh_etal_15}, indicating that the algorithm used in the present work could be a useful tool for the identification of magnetic reconnection sites in numerical simulations. 
Furthermore, these results have important implications for the understanding of fast magnetic reconnection processes, flaring and non-thermal emission in accretion disks and coronae. In particular, as remarked in Section \ref{sec:intro}, recent observations of very rapidly variable, high energy emission associated to compact sources like X-ray binaries and low luminosity AGNs have been interpreted as possibly due to fast reconnection in the coronal regions of these sources \citep[e.g.,][]{dgdp_etal_10b,dgdp_etal_10a, kadowaki_etal_15, singh_etal_15, kushwaha_etal_17}, so that our results offer some support to these studies \citep[see also][about reconnection in Kerr spacetime]{asenjo_comisso_17}. In forthcoming work, we plan to extend the present study exploring the formation of turbulent magnetized accretion disk coronae carrying out global simulations of these systems.  

It is important to stress that the reconnection of the lines has been possible in our ideal MHD simulations because of the underlying numerical resistivity that mimics a small Ohmic resistivity, while the presence of turbulence makes it fast \citep{lazarian_vishiniac_99}. 
\added{In other words, turbulence is the key ingredient to increase the efficiency of the magnetic reconnection rate, which becomes independent of the background resistivity \citep[see also][]{eyink_etal_11, eyink_etal_13, kowal_etal_09, kowal_etal_12, santosLima_etal_10}. The turbulence cascades the magnetic energy down to the kinetic (resistive) scales which are provided by the numerical resistivity in the simulations, but the fast reconnection is controlled by the velocity and length of the turbulence at the injection scales (see eq.\ref{eq:vrec_turb}).}
This is distinct from previous works that explored the effects of an explicit large resistivity  ($\eta$) and viscosity ($\nu$), but still keeping the Prandt number of the order of  unit 
\citep[$P_{m}=\nu/\eta=1$, as in][using non-ideal MHD simulations in shearing-box simulations]{fromang_stone_09}. 

Furthermore, we have dealt with isothermal shearing-box simulations, but an extension of the analysis to a non-isothermal approach could be interesting since thermal diffusivity may play an important role in the dynamics of the system, leading to the expansion of the disk and the development of convection \citep[see][]{bodo_etal_12, bodo_etal_13} that may  also have consequences on the formation of a hot magnetized corona.

\acknowledgments
\textit{Acknowledgments}. The numerical simulations in this work have been performed in the Blue Gene/Q supercomputer supported by the Center for Research Computing (Rice University) and Superintend\^{e}ncia de Tecnologia da Informa\c{c}\~{a}o da Universidade de S\~{a}o Paulo (USP). This work has also made use of the computing facilities of the Laboratory of Astroinformatics (IAG/USP, NAT/Unicsul), whose purchase was made possible by the Brazilian agency FAPESP (grant 2009/54006-4) and the INCT-A; and the cluster of the group of Plasmas and High-Energy Astrophysics (GAPAE), acquired also by the Brazilian agency FAPESP (grant 2013/10559-5). LHSK acknowledges support from the Brazilian agency FAPESP (postdoctoral grant 2016/12320-8) and CNPq (grant 142220/2013-2). EMGDP also acknowledges partial support from the Brazilian agencies FAPESP (grant 2013/10559-5) and CNPq (grant 306598/2009-4). Support from an international cooperation grant between Princeton University and the Universidade de S\~{a}o Paulo is gratefully acknowledged. Also, We would like to thank Kengo Tomida, Zhaohuan Zhu, Grzegorz Kowal, Ji-Ming Shi, and an anonymous referee for useful comments and discussions. 


%\software{astropy \citep{2013A&A...558A..33A},  
%          Cloudy \citep{2013RMxAA..49..137F}, 
%          SExtractor \citep{1996A&AS..117..393B}
%          }

\software{ATHENA code v4.2 \citep{stone_etal_08,stone_etal_10}, VisIt \citep{HPV:VisIt}}

\bibliography{bibliography.bib}

%\begin{thebibliography}{}

%\bibitem[Astropy Collaboration et al.(2013)]{2013A&A...558A..33A} Astropy Collaboration, Robitaille, T.~P., Tollerud, E.~J., et al.\ 2013, \aap, 558, A33 
%\bibitem[Bertin \& Arnouts(1996)]{1996A&AS..117..393B} Bertin, E., \& Arnouts, S.\ 1996, \aaps, 117, 393 
%\bibitem[Corrales(2015)]{2015ApJ...805...23C} Corrales, L.\ 2015, \apj, 805, 23
%\bibitem[Ferland et al.(2013)]{2013RMxAA..49..137F} Ferland, G.~J., Porter, R.~L., van Hoof, P.~A.~M., et al.\ 2013, \rmxaa, 49, 137
%\bibitem[Hanisch \& Biemesderfer(1989)]{1989BAAS...21..780H} Hanisch, R.~J., \& Biemesderfer, C.~D.\ 1989, \baas, 21, 780 
%\bibitem[Lamport(1994)]{lamport94} Lamport, L. 1994, LaTeX: A Document Preparation System, 2nd Edition (Boston, Addison-Wesley Professional)
%\bibitem[Schwarz et al.(2011)]{2011ApJS..197...31S} Schwarz, G.~J., Ness, J.-U., Osborne, J.~P., et al.\ 2011, \apjs, 197, 31  
%\bibitem[Vogt et al.(2014)]{2014ApJ...793..127V} Vogt, F.~P.~A., Dopita, M.~A., Kewley, L.~J., et al.\ 2014, \apj, 793, 127  

%\end{thebibliography}


\listofchanges

\end{document}